\documentclass[journal]{IEEEtran}

\ifCLASSINFOpdf
\else
\fi
\usepackage{multirow}
\usepackage{amsmath, amssymb}
\usepackage{algorithmic}
\usepackage{array}
\usepackage{eulervm}
\usepackage{cleveref}
\ifCLASSOPTIONcompsoc
\usepackage[caption=false,font=normalsize,labelfont=sf,textfont=sf]{subfig}
\else
\usepackage[caption=false,font=footnotesize]{subfig}
\fi
\ifCLASSOPTIONcaptionsoff
\usepackage[nomarkers]{endfloat}
\let\MYoriglatexcaption\caption
\renewcommand{\caption}[2][\relax]{\MYoriglatexcaption[#2]{#2}}
\fi
\usepackage{url}
\usepackage{amsmath,amsfonts}
\usepackage{algorithmic}
\usepackage{algorithm}

\usepackage[normalem]{ulem}
\usepackage{textcomp}
\usepackage{url}
\usepackage{verbatim}
\usepackage{graphicx}
\usepackage{cite}

\usepackage{tikz}
\usetikzlibrary{positioning,spy}

\usepackage{makecell}

\usepackage{makecell}
\usepackage{subfig,graphicx}

\usepackage{mathtools}

\usepackage{xcolor}
\usepackage{soul}
\sethlcolor{yellow} 
\begin{document}
	
	\title{Orthogonal Time Frequency Multiplexing (OTFDM): A Novel Waveform Targeted for IMT-$2030$}

	\author{
		\IEEEauthorblockN{{Koteswara Rao Gudimitla*}\thanks{*Corresponding author}, Sibgath Ali Khan M, Kiran Kuchi}
		\\
		
		\thanks{K.R. Gudimitla and S.A.K. Makandar are with Wisig Networks, Hyderabad, Telangana, India e-mail: (koti, sibgath)@wisig.com}
		\thanks{K. Kuchi is with the Department
			of Electrical Engineering, Indian Institute of Technology Hyderabad, Sangareddy, Telangana, India e-mail: (kkuchi@ee.iith.ac.in).}
		
	}
	\maketitle
	\begin{abstract}
		The rapid evolution of the International Mobile Telecommunications (IMT) landscape has prompted the International Telecommunications Union Working Party $5$D (ITU WP $5$D) to outline the framework for IMT-$2030$ and beyond. This next-generation initiative seeks to meet the diverse demands of future networks, with key objectives including hyper-low latency, enhanced energy efficiency, and robust support for high mobility. Current $5^{th}$ generation ($5$G) technologies employ waveforms like Orthogonal Frequency Division Multiplexing (OFDM) and Discrete Fourier Transform Spread Orthogonal Frequency Division Multiplexing (DFT-s-OFDM). However, these waveforms are insufficient to fully meet the stringent requirements of next-generation
		communication systems. This paper introduces a novel waveform, Orthogonal Time Frequency Division Multiplexing (OTFDM), designed to address the limitations of existing waveforms. OTFDM achieves ultra-low latency by enabling single-shot transmission of data and Reference Signals (RS) within a single symbol. Furthermore, OTFDM supports high mobility with improved resilience to Doppler shifts and enhances power amplifier efficiency through its low Peak-to-Average Power Ratio (PAPR) characteristics. The proposed waveform incorporates advanced signal processing techniques, including time-frequency multiplexing and frequency domain spectrum shaping, to mitigate inter-symbol interference (ISI). These techniques enable accurate per-symbol channel estimation, thus supporting higher-order modulations even at higher user speeds. Extensive simulations validate the efficacy of OTFDM, demonstrating its capability to support user speeds up to $500$ Km/h with minimal RS overhead. This paper explores the technical aspects of OTFDM and discusses its potential implications for the next-generation wireless communication systems.
		
	\end{abstract}
	
	
	\IEEEpeerreviewmaketitle
	\section{Introduction}\label{sec: introduction}
	\IEEEPARstart{T}{he} International Mobile Telecommunications (IMT) landscape is rapidly evolving, marked by the recent ITU WP $5$D recommendation (International Telecommunications Union Working Party $5$D)~\cite{ITU.WPD5IMT2030}. This recommendation outlines the framework and objectives for IMT-$2030$ and beyond. This initiative aims to meet the diverse needs of the networked societies globally, covering a broad range of capabilities for the envisioned usage scenarios.

	Hyper-low latency is one of the key requirements for IMT-$2030$, necessitating a radio access technology that supports
	extremely low latency, e.g., $100 \mu s$. This requires a waveform capable of transmitting data and control signals within a very short Transmission Time Interval (TTI). For hyper-low latency transmission, data must be transmitted in a single symbol or via a single-shot transmission to meet the stringent latency requirements. Several techniques~\cite{Dhiraj, Alphan_Sahin, Arnaud_bouttier} have been explored to achieve single-symbol transmission. However, these methods often compromise the Peak-to-Average Power Ratio (PAPR) of the waveform or introduce significant implementation complexities.

	Another key objective for IMT-$2030$ is to develop radio access technology that is highly energy-efficient. Since the sub-$6$ GHz frequencies, particularly those up to $4$ GHz, are allocated for $5^{th}$ generation ($5$G) technologies, future IMT-$2030$ deployments are planned to utilize higher frequency bands above $6$ GHz, commonly known as upper mid-band frequencies~\cite{ITU.WPD5IMT2030},~\cite{6gfreq}. These higher frequencies experience significant signal propagation losses. Therefore, to enable IMT-2030 to operate effectively in these new frequency bands, it is crucial to adopt techniques that enhance the power and energy efficiency of the power amplifier (PA), increase its output power, and extend the communication range. This requirement necessitates a waveform with substantially higher power/energy efficiency of the PA.

	Additionally, IMT-$2030$ introduces a critical objective to support higher mobility, accommodating user speeds up to $500$ Km/h or higher~\cite{ITU.WPD5IMT2030}. The combination of high mobility and operation at higher carrier frequencies increases Doppler frequency shift, which can degrade symbol decoding performance, particularly for higher-order constellations.
	
	\begin{figure}[t]
		\centering
		\includegraphics[width = \linewidth]{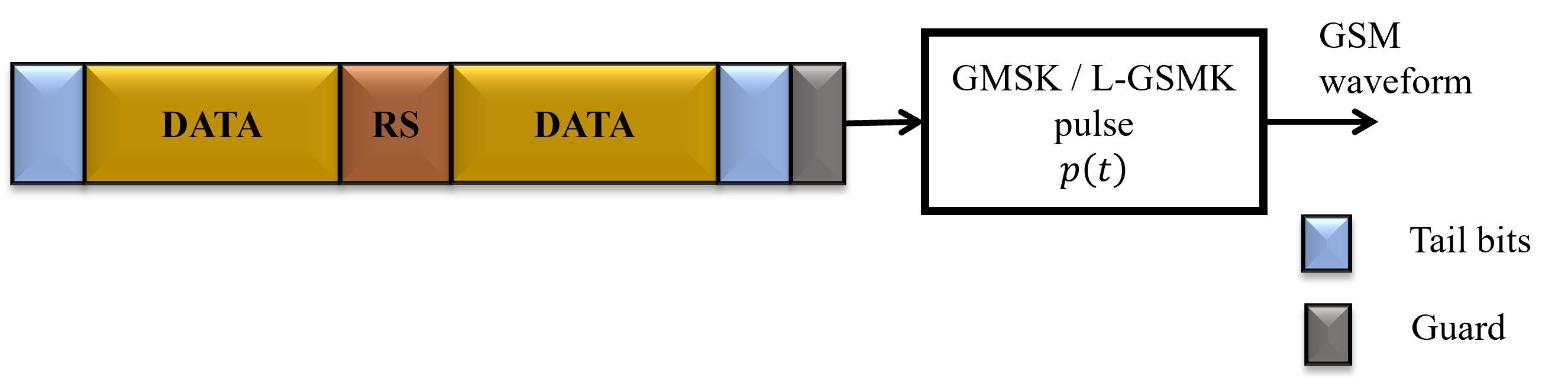}
		\caption{GSM transmit waveform generation.}
		\label{fig: gsm}
	\end{figure}
	
	\begin{figure}[t]
		\centering
		
		\subfloat[OFDM waveform.\label{fig: OFDM}]{%
			\includegraphics[width=\linewidth]{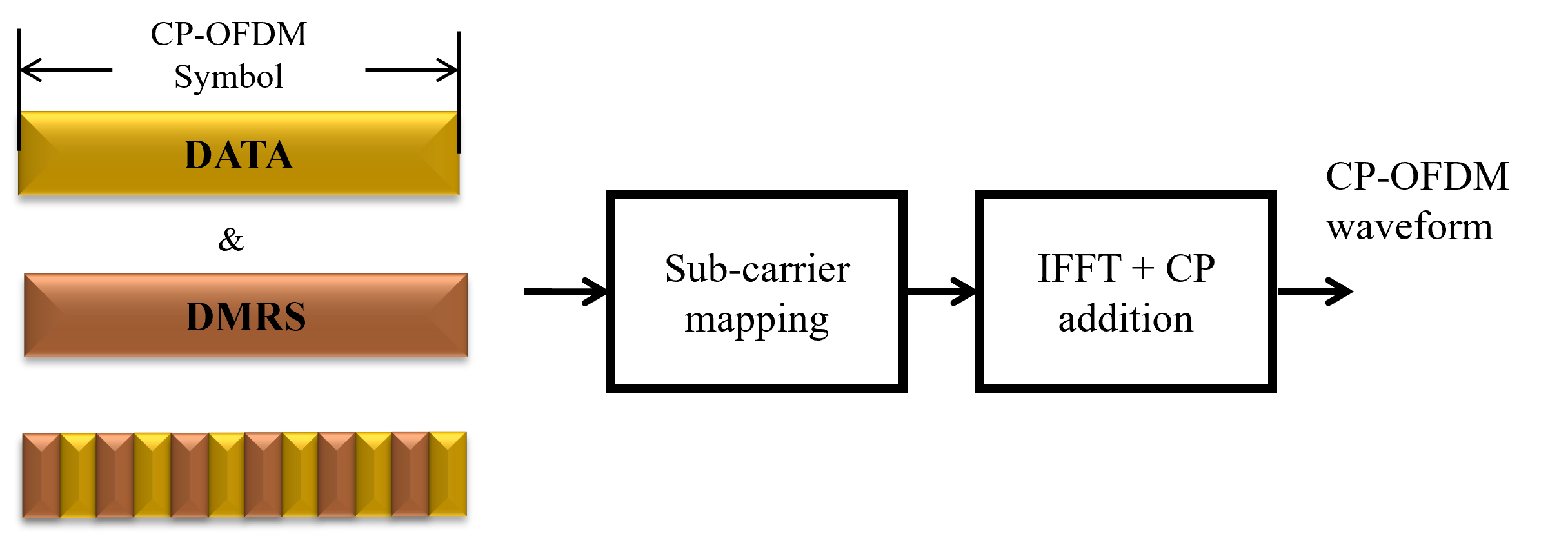}}
		\vfill
		\subfloat[DFT-s-OFDM waveform.\label{fig: Tx_dftsOFDM}]{%
			\includegraphics[width=\linewidth]{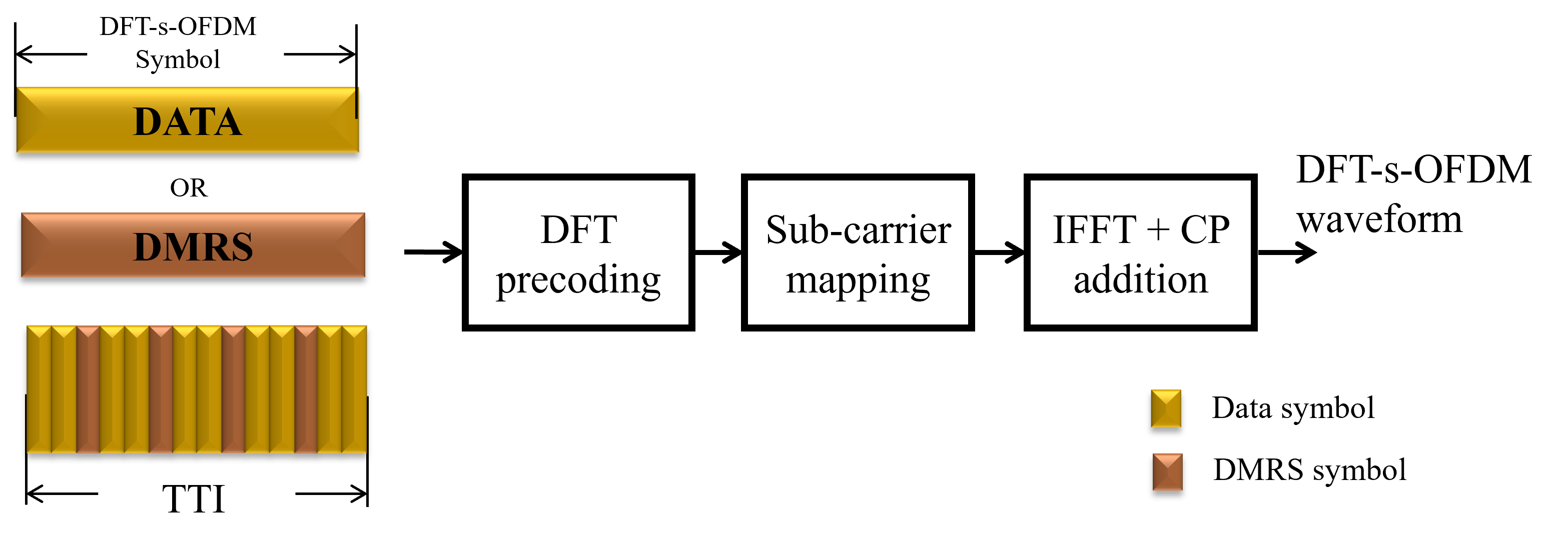}}
		\caption{Waveform generation of CP-OFDM and DFT-s-OFDM.}
		\label{fig: symb_struct} 
	\end{figure}
	\subsection{Traditional waveforms}
	$2^{nd}$ generation ($2$G), Global System for Mobile Communications (GSM), cellular technology used a combination of Time Division Multiple Access (TDMA) and Frequency Division Multiple Access (FDMA) principles, allowing multiple users to share the available resources in both time and frequency. The GSM transmit waveform employs time multiplexing of data and Reference Signal (RS)  within a given transmission slot that is processed using either Gaussian Minimum Shift Keying (GMSK) or a Linearized GMSK (L-GMSK) pulse-shaping filter shown in Fig.~\ref{fig: gsm}. Since GMSK and L-GMSK offer low PAPR, allowing signal transmission near the PA saturation level, thus delivering full power and maximizing coverage. In GSM, the duration of each time slot is $0.577$ms, which includes $26$ RS embedded in between $114$ data symbols, $3$ tail symbols, and $2$ guard symbols at the edges of the slot~\cite{gsmSlotStruct}.

	Orthogonal Frequency Division Multiplexing (OFDM) and Discrete Fourier Transform spread OFDM (DFT-s-OFDM) are the two standard waveforms employed in both $4^{th}$ and $5^{th}$ generation cellular technologies~\cite{3gpp.38.211}. OFDM enables supporting high data rate communications in uplink and downlink (refer to Fig.~\ref{fig: OFDM}). In addition, 5G systems support various subcarrier spacings (SCS), which impact overall transmission latency. For instance, current 5G deployments in Frequency Range 2 (FR2) use a subcarrier spacing of 120 kHz, leading to an OFDM symbol duration of 8.9 microseconds. This short symbol duration enables extremely low-latency transmission. 
	
	While OFDM meets the high data rate and low latency requirements, it suffers from a high Peak-to-Average Power Ratio (PAPR). To mitigate this, PAPR reduction techniques, such as Digital Pre-Distortion (DPD) circuits, are used. However, these methods are complex and costly to implement, leading to lower overall power and energy efficiency in the system. On the other hand, DFT-s-OFDM, a low PAPR waveform, is particularly used in uplink scenarios where coverage is limited. The transmission of DMRS and data takes place on distinct symbols (refer to Fig.~\ref{fig: Tx_dftsOFDM}), and the channel estimates derived from DMRS symbols are either replicated onto the data symbols or undergo time-interpolation.

	Another significant challenge in current  OFDM/DFT-s-OFDM based  $5$G systems is support for high-mobility users. The existing $5$G architecture allows for a maximum of $14$ symbols to be configured for users, with up to $4$ symbols allocated for DMRS transmission shown in Fig.~\ref{fig: Tx_dftsOFDM}. Additionally, DMRS and data are transmitted on separate OFDM symbols. Consequently, when channel estimates derived from the DMRS symbols are directly used to equalize the data, it results in performance degradation. Increasing the number of DMRS symbols or using time-interpolation techniques enhances Doppler mitigation but incurs significant DMRS overhead and implementation complexities ~\cite{NTN_Doppler_compensation, pilot_pattern_design, HighspeedDMRSdesign_journ}.
	
	\begin{figure}[t]
		\centering
		\includegraphics[width=\linewidth]{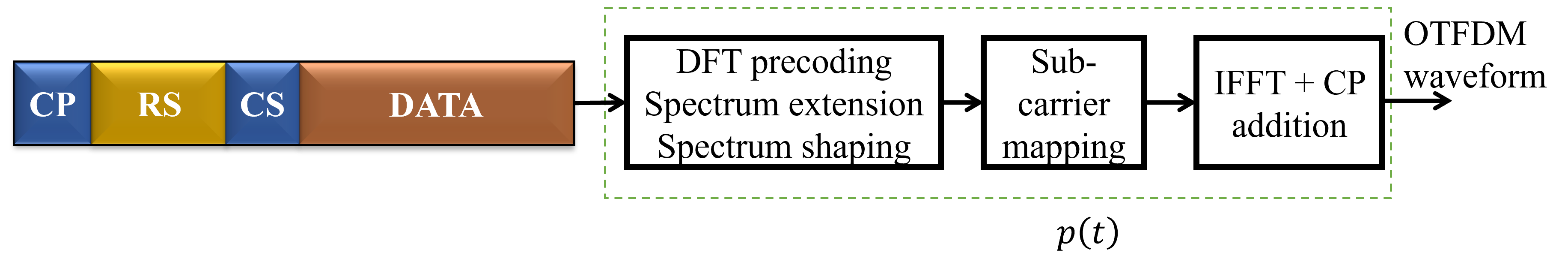}
		\caption{OTFDM transmit waveform generation.}
		\label{fig: OTFDMsymbproc}
	\end{figure}
	
	\begin{figure*}[t]
		\centering
		\subfloat[Transmitter\label{fig: Transmitter}]{%
			\includegraphics[width = \linewidth]{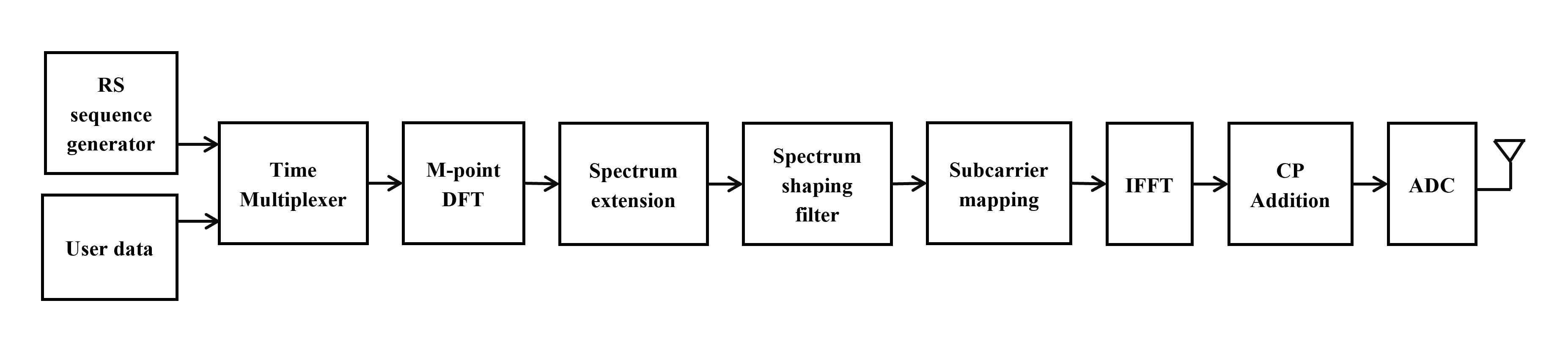}}
		\vfill
		\subfloat[Receiver\label{fig: Receiver}]{%
			\includegraphics[width=\linewidth]{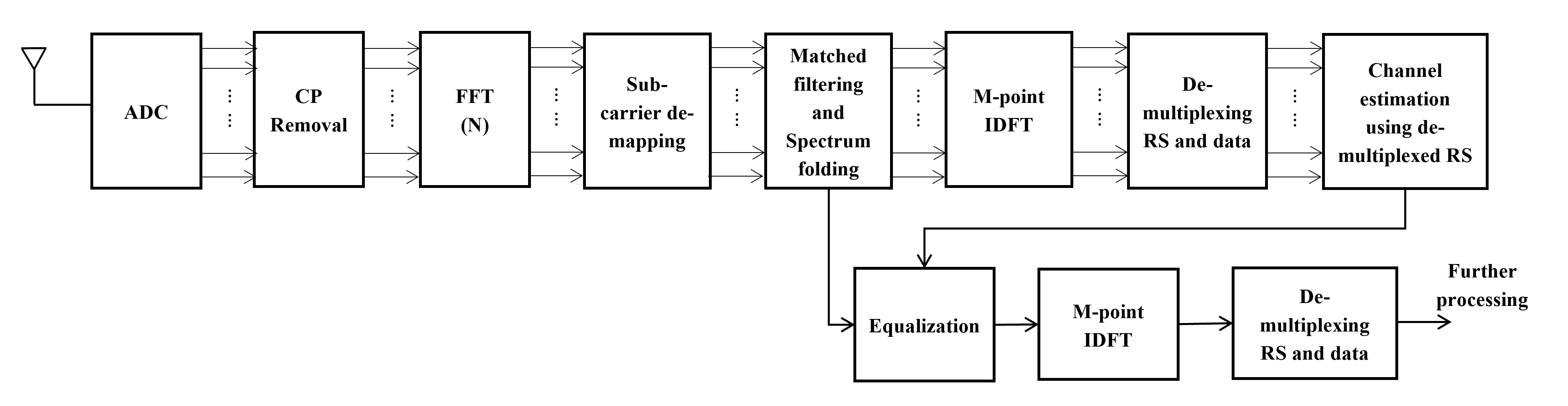}}
		\caption{OTFDM waveform transmitter and receiver architecture.}
		\label{fig: TxRx} 
	\end{figure*}
	
	\subsection{Proposed OTFDM Waveform}
	This paper presents a novel waveform that overcomes the limitations of existing waveforms using a signal-generation method with the following salient features:
	\begin{itemize}
		\item Multiplexing of data and RS within the same symbol along with DFT precoding, excess bandwidth addition, and frequency domain spectrum shaping offers a low PAPR waveform design (refer to Fig.~\ref{fig: OTFDMsymbproc}).
		\item The RS includes its own Cyclic Prefix (CP) and Cyclic Suffix (CS), ensuring efficient Fast Fourier Transform (FFT) based channel estimation.
		\item Simultaneous transmission of RS and data enables the transmission of all user information in a single shot, facilitates self-contained channel estimation, and allows the receiver to instantly demodulate the data.
		\item The proposed waveform improves power and energy efficiency due to its inherent low PAPR characteristics, supports high user mobility and ensures low latency.
	\end{itemize}
	
	The concept of excess bandwidth and shaping were previously studied by $3^{rd}$ Generation Partnership Project ($3$GPP) during the development of both $4$G and $5$G standards for DFT-s-OFDM waveform~\cite{NTTDOC_1, NTTDOC_2, IITH_PDE, fdss_qpsk, LGe, Ericsson, specextconfpap}. In these studies, the primary focus was to reduce the PAPR of Quadrature Phase Shift Keying (QPSK) modulation. 
	
	Further, the time multiplexing  RS  and data was initially introduced in~\cite{IITH_tdoc, IITH_tdoc_2, Samsung_tdoc} for DFT-s-OFDM systems, with the aim of improving cell coverage, particularly for control data transmissions using $\pi/2$-Binary Phase Shift Keying (BPSK) modulation. The authors in~\cite{pre_dft_RS_paper} extended this idea to the uplink data channel and provided a detailed investigation into the benefits of data-RS multiplexing, especially for higher user speeds. However, they noted that while their proposed design supports high user speeds for lower-order modulations, such as QPSK and $\pi/2$-BPSK, it suffers significant performance loss with higher-order modulations.
	
	The primary limitation associated with the RS and data multiplexing methods provided in earlier contributions~\cite{IITH_tdoc, IITH_tdoc_2} is the Inter-Symbol Interference (ISI) caused by the sinc pulse shaping which is inherent in DFT-s-OFDM. When separate symbols are used for RS  and data transmission,  both experience the same ISI  channel,  enabling accurate channel estimation. In contrast,  time multiplexing  RS  with data in the same DFT-s-OFDM symbol does not allow full ISI channel estimation, as the RS duration is shorter than the effective ISI channel length. This results in an irreducible error floor, especially for operational  Signal-to-Noise  Ratios (SNR)  above  $20$-$30$  dB, where the higher-order modulations like $64$-Quadrature Amplitude Modulation (QAM) and $256$-QAM operate.
	
	
	In this paper, we propose a method to use a suitable spectrum shaping (also termed as pulse shaping) to mitigate the adverse effects of ISI. We show that by using appropriate use of signal bandwidth expansion (using additional subcarriers) and a spectrum shaping filter, ISI can be reduced, resulting in an improvement in the quality of channel estimation. This enhancement allows for the support of higher-order modulations, even at increased user speeds.

	The proposed waveform incorporates several key signal-processing operations, shown in~Fig.~\ref{fig: OTFDMsymbproc}. It starts with time multiplexing data and RS to form a multiplexed sequence of length  $M$. This sequence undergoes transform/DFT  precoding using an  $M$ point  DFT,  followed by excess bandwidth addition (or bandwidth expansion) and spectrum shaping. Standard  OFDM  signal generation techniques such as subcarrier mapping, N-point Inverse Fast Fourier Transform (IFFT), and symbol level CP addition are then applied to generate a time-domain waveform. Given that the method involves Time Division Multiplexing of RS and Data, as well as Frequency Division Multiplexing of subcarriers, the waveform is named Orthogonal Time Frequency Multiplexing (OTFDM) waveform. Each symbol in this waveform is referred to as an OTFDM symbol.

	Further, specific reference signals known as ARS can be embedded alongside RS and data. The ARS specifically aids in estimating and tracking the time-varying channel phase (or Doppler) within each symbol. Inclusion of ARS becomes necessary in systems requiring support for very high mobility or carrier phase tracking, as is done in $5$G NR systems~\cite{3gpp.38.211, PTRS_Phasenoise}.
	
	The receiver processing for the OTFDM waveform differs significantly from that of the standard OFDM-based waveforms, involving several distinct operations as shown in Fig.~\ref{fig: Receiver}:
	\begin{itemize} 
		\item Transmit pulse matched filtering and spectrum folding~\cite{FSE, OptimumMeanSquareDFE}.
		\item De-multiplexing of RS and channel estimation.
		\item Channel equalization of the spectrum folded signal~\cite{Jointequainterf, OptimumSTprocess}.
		\item User data extraction and detection, optionally followed by frequency offset estimation and correction using ARS.
	\end{itemize}
	
	The aim of the receiver is to perform channel estimation and equalization using a single snapshot of an OTFDM symbol. The design parameters of RS density, excess bandwidth, and DFT size are carefully selected to eliminate the irreducible error floor caused by the ISI.
	
	Utilizing comprehensive computer simulations, we illustrate the efficiency of the  OTFDM  waveform. Our block error analysis using the standard $3$GPP channel models reveals that OTFDM exhibits significant potential in accommodating extremely high user speeds up to  $500$ Km/h,  even with higher order constellations such as  $256$-QAM. Further,  the simulations also reveal that the proposed waveform offers low RS overheads compared to the existing systems, and our PAPR analysis demonstrates that the OTFDM waveform facilitates low PAPR operation, thereby enhancing PA efficiency.
	
	In the following sections of this paper, we delve into the technical details of OTFDM signal generation and receiver processing. We also present simulation results and discuss its potential applications and implications for future wireless communication systems.

	\textit{Notations:} Vectors are formatted in bold lower-case $\mathbf{x}_t$, scalars are in lower-case without bold ${x}_t (n)$.The subscripts \textcolor{black}{`$t$'} and \textcolor{black}{`$f$'} denote the time domain and the frequency domain, respectively.
	\begin{figure}[t]
		\centering
		\includegraphics[width = 0.7\linewidth]{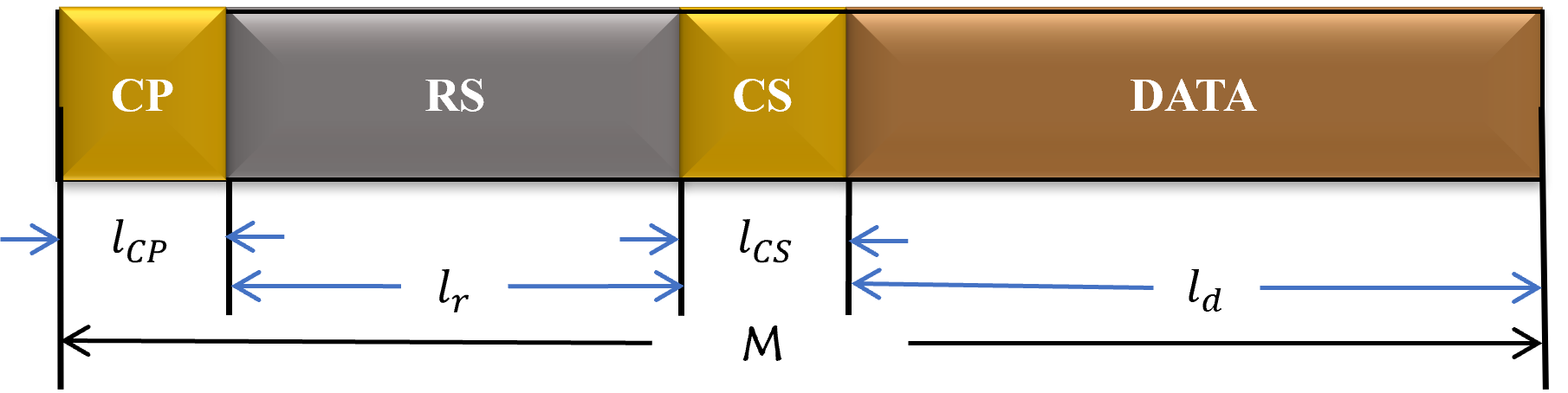}
		\caption{OTFDM symbol representing multiplexed data and Reference Signals.}
		\label{fig: symbol_wo_pts}
	\end{figure}
	

	\section{Transmitter}\label{sec: transmitter}
	In this section, we outline the transmitter architecture for the proposed OTFDM waveform. The block diagram depicting the proposed transmitter is shown in Fig.~\ref{fig: Transmitter}. The core concept of the proposed OTFDM waveform entails time-multiplexing RS and data within a single OTFDM symbol, as illustrated in Fig.~\ref{fig: symbol_wo_pts}. This time-multiplexed sequence is then subjected to DFT precoding, followed by the addition of excess bandwidth and spectrum shaping, before being processed by the OFDM modulator. Due to spectrum shaping, the modulation alphabets of data and RS symbols experience mutual interference, leading to inter-symbol interference (ISI), which can degrade channel estimation performance. To mitigate ISI and enable accurate channel estimation, a guard interval, in the form of a Cyclic Prefix (CP) and a Cyclic Suffix (CS), is appended to the beginning and end of the RS, respectively. Additionally, incorporating CP and CS ensures that a portion of the received RS signal samples is represented as a circular convolution of the RS sequence and the wireless channel. This property holds true as long as the wireless channel length remains time-limited to the duration of the RS.

	Consider $\mathbf{r}_t$  to be the $l_r$ length RS sequence, where, $\mathbf{r}_t \in \mathbb{C}^{l_r\times 1}$. RS is appended with CP and CS, where the last $l_{\scalebox{0.5}{CP}}$ samples of RS form the CP, and beginning $l_{\scalebox{0.5}{CS}}$ samples form the CS. $\mathbf{d}_t$ represents a block of $l_d$ data samples, $\mathbf{x}_t$ represents the data-RS multiplexed symbol such that
	\begin{equation}\label{eq: Txtimesymbnovel_CP}
		{{x}}_t (n) =  \begin{cases*}
			r_t((n-l_{{\scalebox{0.5} {${CS}$}}})~ mod~ l_{{\scalebox{0.5} {${r}$}}}) &, $n = \{0, 1, \dots$, $l_{{\scalebox{0.5} {${RS}$}}}-1\}$, \\
			d_t(n-l_{{\scalebox{0.5} {${RS}$}}}) &, $n = \{l_{{\scalebox{0.5} {${RS}$}}},\dots, M-1\}$, \\
		\end{cases*}
	\end{equation}
	here, $M$ is the length of the symbol (or allocation size), which can be expressed as $M = l_d + l_{RS}$, where $l_{RS} = l_{CP}+l_r+l_{CS}$ is the length of complete RS block (An alternative for data-RS multiplexed symbol is presented in Appendix~\ref{app: onesided}). The multiplexed symbol $\mathbf{{x}}_t$ is subjected to an $M$-point DFT. The resulting DFT precoded OTFDM symbol can be expressed as:
	
	\begin{equation} \label{eq: Precoded sequence}
		x_f(k) = \sum_{n = 0}^{M-1} x_t(n) e^{\frac{-j2\pi kn}{M}},
	\end{equation}
	where, $k$ denotes the subcarrier indices, $k = \{0, 1, 2,\dots M-1\}$. The DFT precoded output $x_f(k)$ is periodic with period $M$, i.e., $x_f(M+k) = x_f(k)$. 
	\begin{figure}
		\centering
		\includegraphics[width = \linewidth]{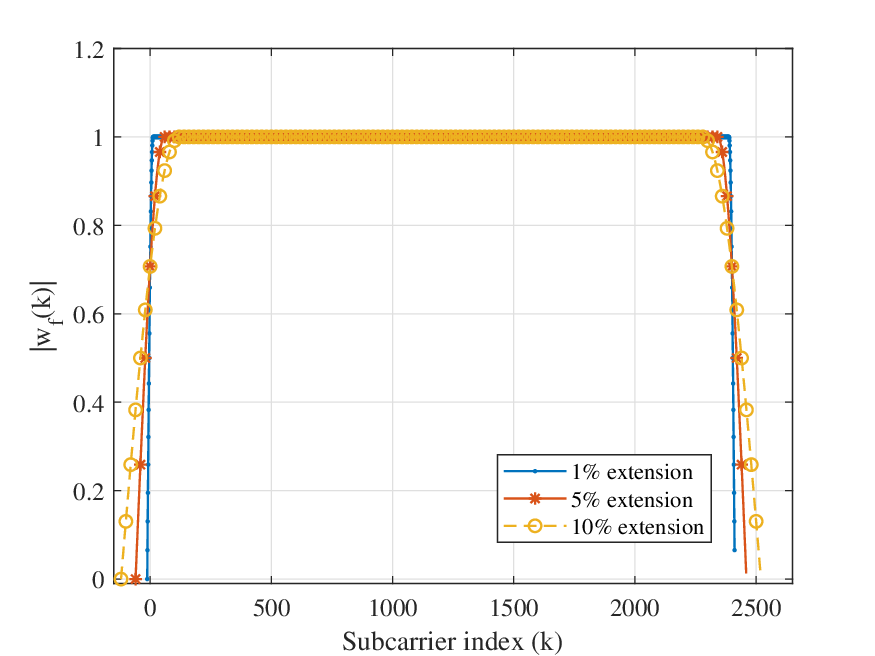}
		\caption{Spectrum of SQRC pulse with different extension factors.}
		\label{fig: sqrcpulse}
	\end{figure}
	The DFT precoded multiplexed symbol $x_f(k)$ undergoes an $\alpha$-fold periodic extension (where $\alpha >1$) to generate a cyclically extended precoded sequence $x_e(k')$. 
	\begin{equation}\label{eq: extension}
		x_e(k{'}) = \begin{cases*}
			x_f(k{'}~mod~M)&\hspace{-11pt}, $k' = \{-\gamma, \dots, 0, \dots, M+\gamma-1\}$,\\
			0 &\hspace{-11pt}, otherwise,
		\end{cases*}
	\end{equation}
	here, $\gamma$ is the number of subcarriers (termed as excess bandwidth) used on each side of the precoded symbol, and the length of the extended sequence is $M+2\gamma$ with $\alpha = 1+\frac{2\gamma}{M}$, $\frac{2\gamma}{M}\times 100$, termed as extension factor, is percentage of the excess bandwidth. Subsequently, the cyclically extended sequence $x_e(k')$ is filtered at the subcarrier level using a Nyquist-class of frequency domain spectrum shaping filter ($\mathbf{w}_f$) to yield a spectrally shaped sequence. For the proposed waveform, we consider a discrete Square-Root-Raised-Cosine (SQRC) filter that is obtained by sampling the continuous frequency SQRC filter:
	\begin{equation}\label{eq: shaping}
		x_s(k') = w_f(k') x_e(k').
	\end{equation} 
	The frequency domain response of the SQRC filter is given by 
	\begin{equation}\label{eq: sqrc}
		{{w}}_f(k') =  \begin{cases*}
			1&\hspace{-11pt}, \scalebox{0.75}{$k' = \{\gamma, \dots, M-\gamma-1\}$}, \\
			\sqrt{\frac{1}{2}\bigl(1+cos(\frac{\pi(-k'+\gamma)}{2\gamma})\bigl)}&\hspace{-11pt}, \scalebox{0.75}{$k' = \{-\gamma, 1, \dots, \gamma-1\}$}, \\
			\hspace{-5pt}\sqrt{\frac{1}{2}\bigl(1+cos(\frac{\pi(k'-M+\gamma)}{2\gamma})\bigl)}&\hspace{-12pt}, \scalebox{0.75}{$k' = \{M-\gamma, \dots, M+\gamma-1\}$}, \\
			0 &\hspace{-10pt}, otherwise,
		\end{cases*}
	\end{equation}
	and Fig.~\ref{fig: sqrcpulse} shows the amplitude spectrum of SQRC pulse with different extension factors. The spectrally shaped RS-data sequence is mapped to a contiguous set of $M+2\gamma$ allocated subcarriers in the frequency domain. It is important to note that filtered sub-carriers can be mapped anywhere within the overall system bandwidth, depending on the resource allocation. However, for the sake of convenience, we assume that the sequence is symmetrically mapped around subcarrier zero, as shown below.
	\begin{equation}\label{eq: Mapping}
		x_m(k) = \begin{cases*}
			x_s (k+\frac{M}{2})&\hspace{-11pt}, $k=\{-\frac{M}{2}-\gamma,\dots, 0,\dots \frac{M}{2}+\gamma-1\}$,\\
			0 &\hspace{-11pt}, ${-\frac{N}{2} \leq k < -\frac{M}{2}-\gamma}$,\\
			0 &\hspace{-11pt}, $\frac{N}{2}-1 \geq k>\frac{M}{2}-\gamma$.
		\end{cases*}
	\end{equation}
	
	\begin{figure}
		\centering
		\includegraphics[width = 0.9\linewidth]{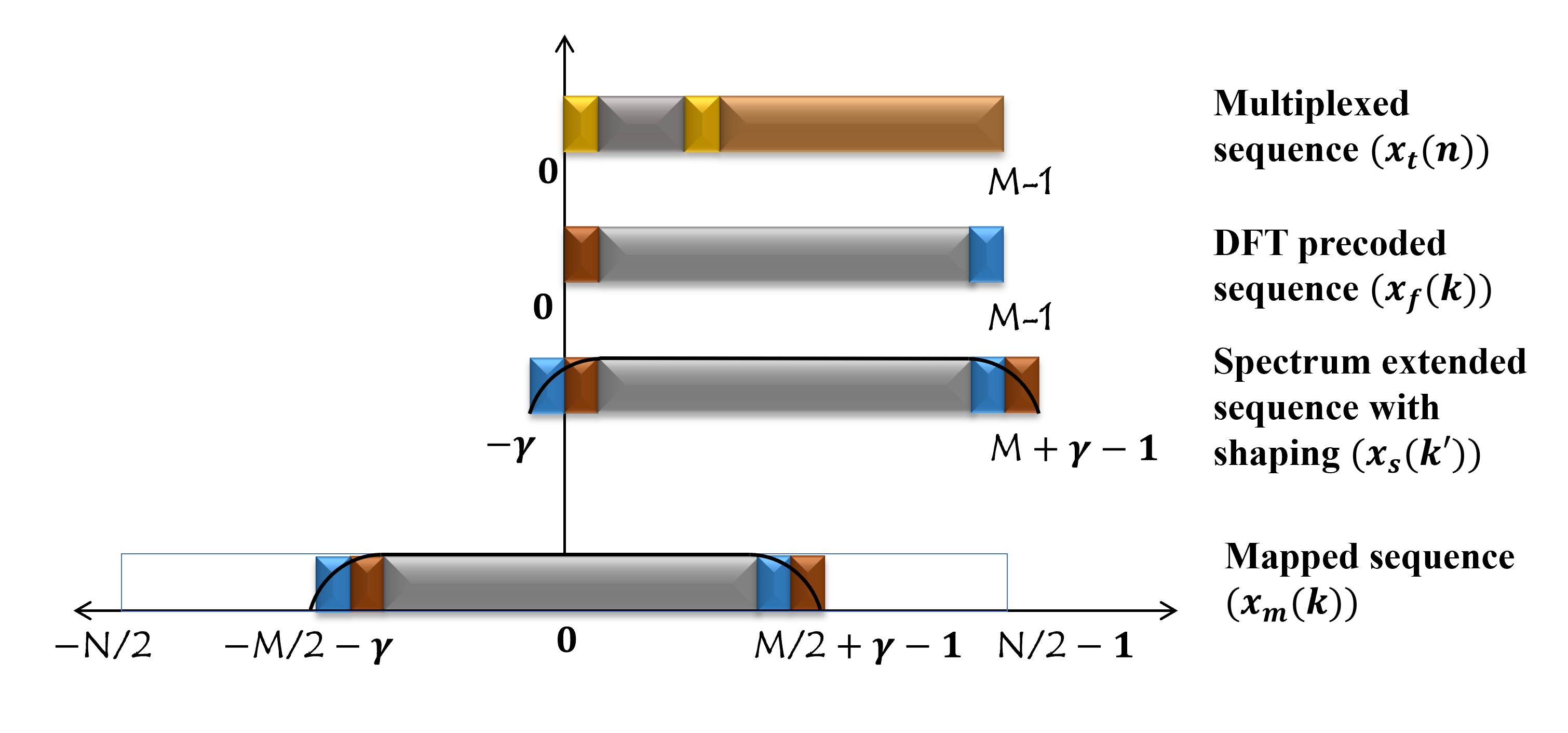}
		\caption{Summary of signal processing steps needed for OTFDM waveform generation.}
		\label{fig: symbproc}
	\end{figure}
	The subcarrier-mapped sequence undergoes standard OFDM operations, which entail an $N$-point IFFT along with the addition of $N_{{\scalebox{0.5} {{CP}}}}$ samples of a symbol level CP. This process generates a time-domain baseband signal $s(t)$ referred to as an OTFDM waveform, as illustrated below.
	\begin{equation}\label{eq: finalTxwaveform}
		s(t) = \sum_{k = -\frac{N}{2}}^{\frac{N}{2}-1} x_m\big(k\big) e^{j2 \pi k \Delta f (t-T_{{\scalebox{0.5} {${CP}$}}})}; t \in [0,(T_{{\scalebox{0.5} {${OTFDM}$}}}+T_{{\scalebox{0.5} {${CP}$}}})],
	\end{equation}
	where $N$ is an appropriately chosen IFFT size, $\Delta f = \frac{1}{T_{{\scalebox{0.5} {{OTFDM}}}}}$ is the subcarrier spacing, $T_{{\scalebox{0.5} {{OTFDM}}}}$ denotes  the signal portion of the OTFDM waveform. $T_{{\scalebox{0.5} {{CP}}}}$ is the duration of the symbol level CP. The figure representing the steps involving OTFDM waveform generation is shown in Fig.~\ref{fig: symbproc}. Further, using \eqref{eq: Txtimesymbnovel_CP}-\eqref{eq: Mapping}, the time domain baseband signal can be represented as a modulated signal, 
	
	\begin{figure}
		\centering
		\includegraphics[width = 0.9\linewidth]{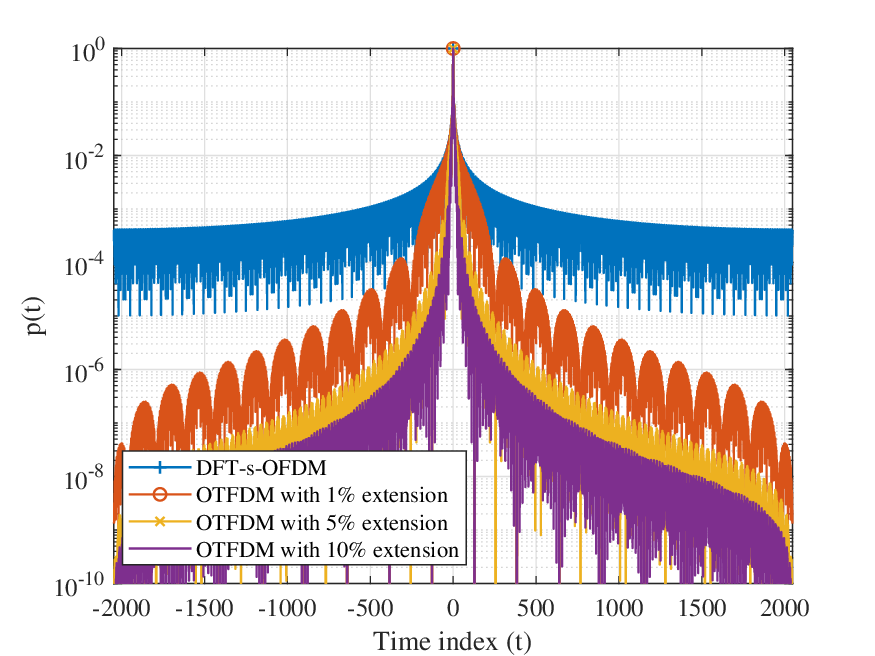}
		\caption{Effective pulse ($p(t)$) with different excess bandwidth factors.}
		\label{fig: effectivepulse_Tx}
	\end{figure}
	
	\begin{align}\label{eq: Alternative Tx expn}
		s(t) = \sum_{n = 0}^{M-1} x_t(n) p(t-nT_s),
	\end{align}
	here, $T_s = \frac{T_{OTFDM}}{M}$, $p(t)$ is the effective pulse acting on the multiplexed symbol~\eqref{eq: Txtimesymbnovel_CP}, and can be shown as
	\begin{equation}
		p(t) = \sum_{k = -N/2}^{N/2-1} w_f(k) e^{-j2\pi k\Delta f(t-T_{CP})}.
	\end{equation}
	Given that the spectrum shaping filter $w_f(k)$ is a Nyquist filter, expanding its bandwidth enhances the localization of $p(t)$ in the time domain, which further reduces the side lobes~\cite{Proakis, nyquistcriterion}. Fig.~\ref{fig: effectivepulse_Tx} illustrates the behavior of the effective impulse response across different extension factors. It can be observed that by increasing the excess bandwidth, the tails of the impulse response decay much more rapidly, significantly reducing channel energy leakage and thereby enhancing the accuracy of channel estimation.
	
	\begin{figure}[t]
		\centering
		\includegraphics[width = 0.7\linewidth]{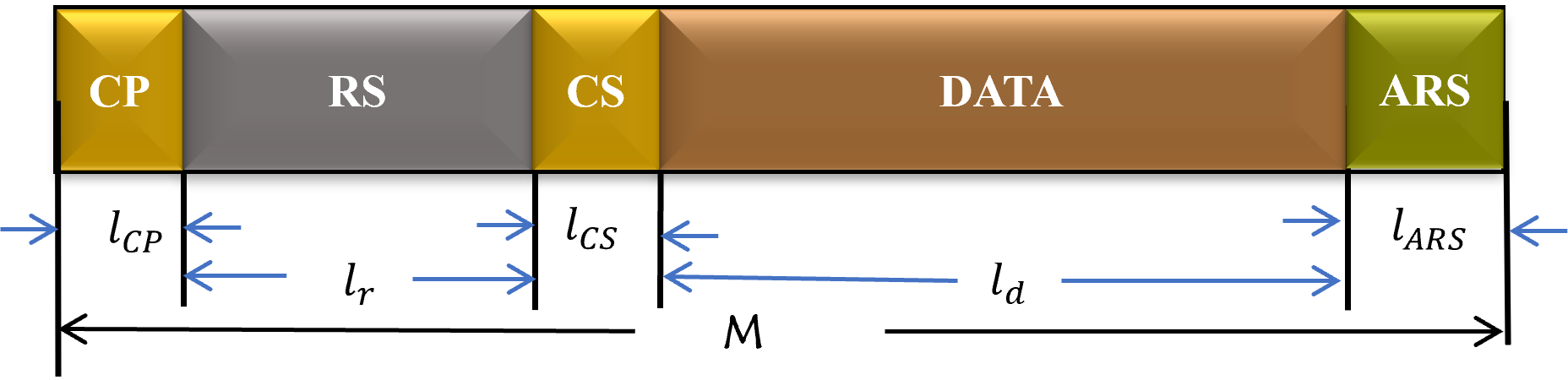}
		\caption{OTFDM symbol representing multiplexed data and Reference Signals with Additional Reference Signals.}
		\label{fig: symbol_w_pts}
	\end{figure}
	
	\subsection{Need for Additional Reference Signals (ARS)}
	When systems operate at high carrier frequencies, such as millimeter-wave frequencies ( $> 28$ GHz), the crystal oscillator frequency can drift over time, resulting in a frequency offset. Similarly, in high-speed train environments, the channel between the transmitter and receiver undergoes rapid variations due to the train's high-speed movement. In these scenarios, the receiver must perform residual phase offset correction in addition to channel estimation and equalization. If the phase offset is not corrected, it can lead to severe performance degradation, particularly for higher-order constellations.

	ARS signals are optionally transmitted alongside RS and data within an OTFDM symbol to estimate the residual phase offset. In this study, we consider placing $l_{ARS}$ length ARS at the tail end of the OTFDM symbol such that the phase growth from RS samples is high and considerably good for phase estimation. ARS multiplexed with data and RS is as illustrated in Fig.~\ref{fig: symbol_w_pts}, and the multiplexed symbol can be represented by
	
	\begin{equation}\label{eq: Txtimesymbnovel_ARS}
		{{x}}_t (n) =  \begin{cases*}
			r_t((n-l_{{\scalebox{0.5} {${CS}$}}})~ mod~ l_{{\scalebox{0.5} {${r}$}}}) &, $n = \{0, 1, \dots$, $l_{{\scalebox{0.5} {${RS}$}}}-1\}$, \\
			d_t(n-l_{{\scalebox{0.5} {${RS}$}}}) &, $n = \{l_{{\scalebox{0.5} {${RS}$}}},\dots, l_d +l_{{\scalebox{0.5} {${RS}$}}}-1\}$, \\
			a_t(n-l_d -l_{{\scalebox{0.5} {${RS}$}}})&, $n = \{l_d +l_{{\scalebox{0.5} {${RS}$}}},\dots, M-1\}$.\\
		\end{cases*}
	\end{equation}
	Alternatively, periodically spaced pilots within the OTFDM symbol can be used to track and compensate for channel phase variations over time. Including ARS has proven effective in mitigating the time variations caused by high-speed trains. The performance improvements due to ARS are discussed in detail in Results Section~\ref{sec: results}.

	\section{Receiver}\label{sec: receiver}
	This section outlines the essential processing blocks involved in equalizing the OTFDM waveform. The functional block diagram of the receiver is depicted in Fig.~\ref{fig: Receiver}. Notably, the OTFDM receiver processing differs substantially from the standard OFDM receiver and includes the following key operations.
	
	\begin{itemize}
		\item Transmit pulse matched filtering and Spectrum folding ~\cite{FSE, OptimumMeanSquareDFE}.
		\item De-multiplexing of RS from OTFDM signal and performing channel estimation.
		\item Channel Equalization of the spectrum folded signal~\cite{Jointequainterf, OptimumSTprocess}.
		\item User data extraction and detection, optionally followed by frequency offset estimation and correction using ARS.
	\end{itemize}
	
	The transmit OTFDM signal goes through a wireless channel $h_{prop}(t)$ and complex-valued Additive White Gaussian Noise (AWGN) $v(t)$. The received OTFDM signal is given by
	\begin{align}\label{eq: received signal}
		y(t) = s(t) \star h_{prop}(t) + v(t)
	\end{align}
	where $\star$ denotes linear convolution operation. The received OTFDM signal first undergoes receiver front-end processing, which comprises the Analog-to-Digital-Converter (ADC), symbol level CP removal (post IFFT CP), N-point FFT, and subcarrier demapping. The demapped OTFDM signal is represented as:
	\begin{equation}\label{eq: demapdata}
		y_s(k_1)= h_s(k_1) x_s(k_1)+v_s(k_1),
	\end{equation}   
	here, $k_1 =  \{-\gamma, -\gamma+1, \dots, M+\gamma-1 \}$, $v_s(k_1)$ is complex-valued AWGN with a variance of $\sigma^2$, and $h_s$ denotes the overall channel impulse response in the frequency domain that includes the spectrum shaping filter and the propagation channel.
	
	\subsection{Obtaining $M$ subcarriers from $M+2\gamma$ subcarriers}
	
	Since OTFDM signal generation includes bandwidth expansion, the length of the demapped sequence ($M+2\gamma$) exceeds the actual length of the multiplexed Data-RS. Therefore, we adopt an approach that includes matched filtering and spectrum folding to obtain $M$ subcarriers as represented below~\cite{FSE, OptimumMeanSquareDFE},
	\begin{equation}\label{eq: matchedfilteredoutput}
		y_f(k) = \sum_{p = -1}^{1} w_f(k+pM-\gamma) y_s(k+pM-\gamma)
	\end{equation}	
	here, $k = \{0, 1, 2, \dots, M-1\}$. Note that matched filtering maximizes the SNR, and the spectrum folding is essential to obtain symbol rate samples, i.e., $M$ samples per symbol. For matched filtering, a Nyquist-class filter is applied to the demapped OTFDM signal. It is important to note that the same filter was used for OTFDM signal generation at the transmitter. As the excess bandwidth is assumed to be less than $100$ percent, the folded spectrum will be flat over the bandwidth of interest~\cite{Proakis, OptimumMeanSquareDFE}, i.e., $M$ subcarriers, that is: 
	\begin{equation}
		\sum_{p = -1}^{1} |w_f(k+pM-\gamma)|^2 = 1
	\end{equation}
	where, $k = \{0, 1, 2, \dots, M-1\}$. It is important to note that matched filtering and spectrum folding operations are equivalent to symbol rate sampling used in traditional equalizers for single carrier systems~\cite{FSE, FaterthanNyquist, nyquistcriterion}. 
	
	When the frequency spectrum of the wireless channel is flat, zero ISI can be perfectly ensured. However, in practical scenarios, two situations may arise:
	\begin{itemize}
		\item In Line-of-Sight (LOS) systems, the receiver encounters finite sampling timing errors, leading to ISI that requires estimation and equalization.
		\item In dispersive channels, due to the absence of an optimal sampling timing reference, the receiver inevitably experiences ISI. 
	\end{itemize}
	To counter the dispersion, channel estimation is performed, and the resultant channel estimates are used to equalize the $M$-length spectrum folded OTFDM symbol in~\eqref{eq: matchedfilteredoutput}. Subsequently, the equalized signal is subjected to an $M$-length Inverse Discrete Fourier Transform (IDFT) to obtain the desired data samples, which are then fed to bit processing modules. 
	
	\begin{figure*}[t]
		\centering
		\subfloat[Wireless channel $h_{prop}(t)$ \label{fig: TDLC100} ]{\includegraphics[width = 0.33\linewidth]{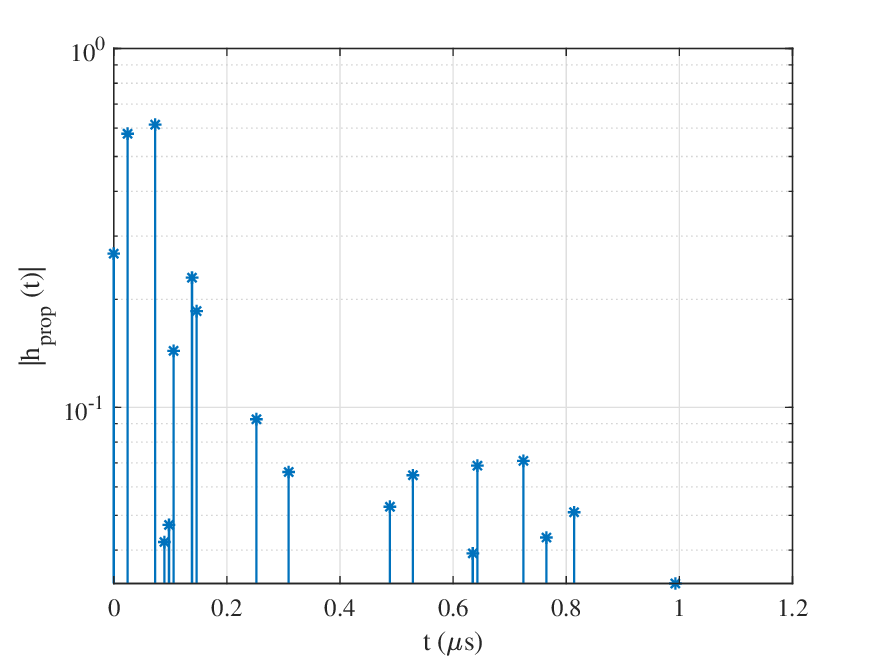}}
		\subfloat[Effect of excess bandwidth and shaping on the channel $h_t(n)$ \label{fig: chan_localization} ]{\includegraphics[width = 0.33\linewidth]{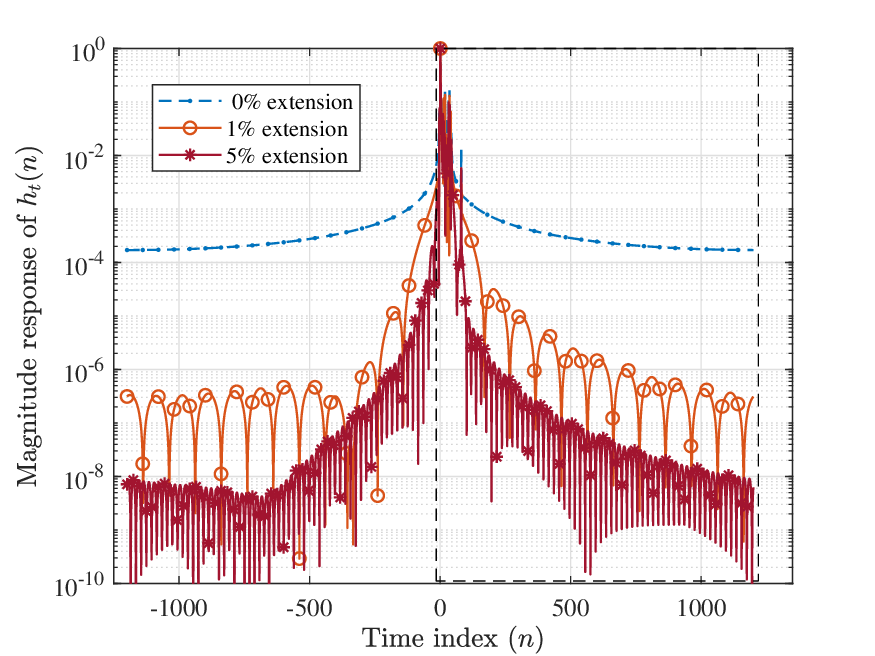}}
		\subfloat[Tail lobes decaying as a function of excess bandwidth factor \label{fig: tails} ]{\includegraphics[width = 0.33\linewidth]{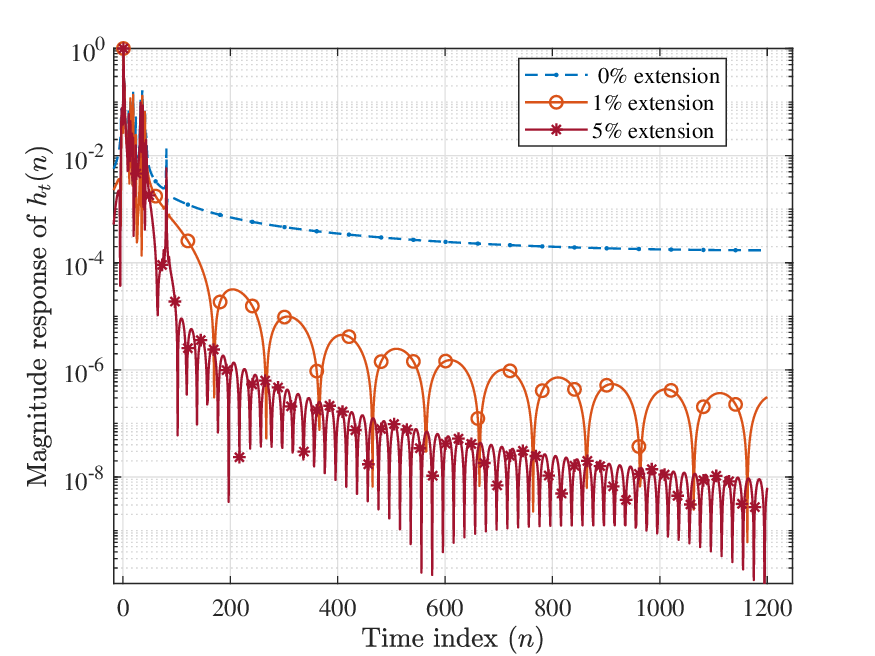}}
		\caption{Effect of excess bandwidth and shaping on TDL-C channel with a maximum delay spread of $1 \mu s$.}
		\label{fig: chan_localization_diff_exs_fac} 
	\end{figure*}
	\subsection{Channel estimation using RS} \label{subsec: Chan esti}
	In the proposed OTFDM waveform, data and RS are multiplexed in the same symbol and are jointly precoded using an $M$-point DFT. Hence, it is crucial to note that although data and RS are temporally separated, the DFT output represents a composite mixture of data and RS samples. Consequently, RS are not readily available for channel estimation after filtering and spectrum folding. Hence, the spectrum folded OTFDM symbol in~\eqref{eq: matchedfilteredoutput} is transformed to the time domain through an IDFT operation. This operation results in the reconstruction of a time-domain symbol, which comprises multiplexed RS and data along with RS CP and RS CS, as illustrated below.
	
	\begin{equation}\label{eq: RxedIDFT}
		y_t(n) = \frac{1}{M}  \sum_{k=0}^{M-1} y_f(k) e^{\frac{j2\pi n k}{M}};  n={0, 1, 2, \dots, M-1},
	\end{equation}
	\begin{equation}\label{eq:Rxtimesymb1_modifiedRS1}
		\mathbf{{y}}_t = \begin{bmatrix}
			\mathbf{y}_{t, {\scalebox{0.5} {${CP}$}}} \\ \mathbf{y}_{t, {\scalebox{0.5} {${RS}$}}} \\ \mathbf{y}_{t, { {\scalebox{0.5} {${CS}$}}}} \\ \mathbf{y}_{t, d}
		\end{bmatrix}.
	\end{equation}
	The RS is appended with CP and CS primarily for ISI mitigation and to bring in circular convolution between RS and the ISI channel (to be estimated). Hence, excluding the CP and CS samples, only the $l_r$ length RS samples are extracted from $y_{t}(n)$ for channel estimation. In this work, we apply the frequency domain channel estimation method based on the Least Squares (LS) method. First, the $l_r$ length RS samples are transformed to frequency domain through $l_r$ point DFT, as:
	\begin{equation}\label{eq: Freq domain rxed RS}
		{y}_{f, {\scalebox{0.5} {${RS}$}}}(k)= \sum_{n=-\frac{l_r}{2}}^{\frac{l_r}{2}-1} {y}_{t, {\scalebox{0.5} {${RS}$}}}\bigl(n+\frac{l_r}{2} \bigl) e^{\frac{-j2\pi nk}{l_r}}.
	\end{equation}
	Here, $k = \{-\frac{l_r}{2},\dots, 0, 1,\dots, \frac{l_r}{2}-1\}$. LS method is applied between ${y}_{f, {\scalebox{0.5} {${RS}$}}}(k)$ and ${{r}_{f}(k)}$ to estimate joint frequency response of the wireless channel and spectrum folded filter as shown below
	\begin{equation}\label{eq: LS estimation}
		{h}_{f, {\scalebox{0.5} {${RS}$}}}(k) = \frac {{y}_{f, {\scalebox{0.5} {${RS}$}}}(k)} {{r}_{f}(k)}; ~|r_f(k)|> 0,
	\end{equation} 
	here, ${r}_{f}(k)$ is the $l_r$ length frequency domain RS sequence. However, to equalize the spectrum folded symbol $y_f(k)$, $M$ length frequency domain channel estimates are needed. This requires interpolating $l_r$ length channel to an $M$-length channel. Hence, in our work, we apply an $l_r$ point IDFT to~\eqref{eq: LS estimation}, which results in time domain channel impulse response.
	\begin{equation}\label{eq: RS size time domain chan}
		{h}_{t, {\scalebox{0.5} {${RS}$}}}(n) = \frac{1}{{l_r}}\sum_{k = -\frac{l_r}{2}}^{\frac{l_r}{2}-1} {h}_{f, {\scalebox{0.5} {${RS}$}}}(k) e^{\frac{j2\pi nk}{l_r}}.
	\end{equation}
	Since the time domain channel impulse response of interest is usually limited to a certain length, a time domain window  $w_h(n)$ is applied on~\eqref{eq: RS size time domain chan} to obtain the desired time domain impulse response. This operation also limits the additive white noise power~\cite{noisefiltering, on_CE_OFDM}. Thus, we have:
	\begin{equation}\label{eq: filtered time domain estimates}
		{h}_{w, {\scalebox{0.5} {${RS}$}}}(n)  = w_h(n) {h}_{t, {\scalebox{0.5} {${RS}$}}}(n).
	\end{equation}
	To obtain frequency domain channel estimates of length $M$, an $M$-point DFT is applied on the impulse response ${h}_{w, {\scalebox{0.5} {{RS}}}}$ after including an appropriate number of zeros. We get:
	\begin{equation}\label{eq: M_len_chan}
		\hat{h}_f(k)=\sum_{n=0}^{M-1} {h}_{w, {\scalebox{0.5} {${RS}$}}}(n) e^{\frac{-j2\pi nk}{M}}
	\end{equation}
	here,  $k={0, 1, 2, \dots, M-1}$, and $ \hat{h}_f(k)$.
	
	\subsection{Equalization}
	The $M$-point estimated channel is used to equalize the spectrum folded signal using a Minimum Mean Squared Error (MMSE) equalizer~\cite{Cioffi}. The spectrum folded output signal on each subcarrier is equalized using  $w(k)$ as:
	\begin{equation}
		\hat{x}_f(k) = w(k) y_f(k),
	\end{equation}	
	and $w(k)$ is:
	\begin{equation}
		w(k)=\frac{(\hat{h}_f(k))^\dagger}{|\hat{h}_f(k)|^2+ \sigma^2 },
	\end{equation}	
	where, $k = \{0, 1, 2, \dots, M-1\}$. It is noteworthy that the noise variance ($\sigma ^2$) will remain the same after filtering and spectrum folding operations when the Nyquist class of filters are used. To obtain the equalized data corresponding to the transmitted multiplexed data, an IDFT of size $M$ is performed on $\hat{x}_f(k)$, which is processed further for demodulation and decoding.

	\subsection{Impact of excess bandwidth and shaping on channel $h_t(n)$}
	In this subsection, we highlight the role of excess bandwidth and shaping in the generation of the OTFDM waveform. When RS and data are time-multiplexed within the same symbol, the ISI channel has substantial tail energy that spreads across the time samples, ultimately degrading the quality of channel estimates. For the proposed OTFDM waveform, by employing both excess bandwidth and spectrum shaping, the tails of the effective channel can be greatly reduced, as demonstrated in Fig.~\ref{fig: chan_localization_diff_exs_fac}. The impulse response of a typical Tapped Delay Line-C (TDL-C) propagation channel is shown in Fig.~\ref{fig: TDLC100}, whereas Figs.~\ref{fig: chan_localization},~\ref{fig: tails} depicts effective channel impulse response observed at the receiver under different extension factors. It can be observed from the figures that without extension and shaping, sinc interpolation causes the channel energy to spread across the entire allocation size $M$, which is $2400$ samples in this case. However, employing the SQRC pulse shaping along with receiver-matched filtering significantly reduces channel energy leakage, and the impulse response tails decay much faster with larger excess bandwidth values.
	
	\subsubsection{Impact of excess bandwidth and shaping on RS length, CP, CS Size}\label{subsec: RS size selection}
	In the conventional DFT-s-OFDM waveform, the RS length should be equal to the symbol length to reliably estimate the impulse response of the propagation channel. However, for the proposed OTFDM waveform, by utilizing excess bandwidth and spectrum shaping, it is possible to localize the channel energy. Consequently, the effective length of the impulse response becomes significantly less than the symbol length $M$. This assumption is valid when $M$ is large. Hence, with the use of spectrum shaping and excess bandwidth, the RS length required to estimate these channel taps is reduced, and hence, the sizes of corresponding CP and CS.
	
	\begin{figure}
		\centering
		\includegraphics[width = \linewidth]{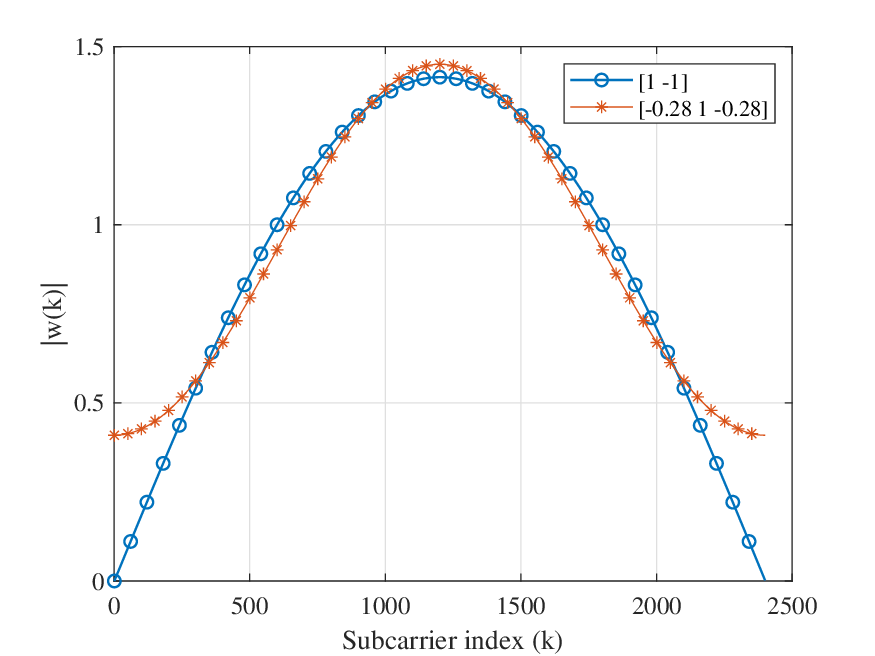}
		\caption{Magnitude spectrum of $2$-Tap and $3$-Tap filters.}
		\label{fig: 2tapfil}
	\end{figure}
	\subsection{ $\pi/2$-BPSK data transmissions}\label{Subsec: pi2data}
	$\pi/2$-BPSK modulation is introduced in $5$G-NR to support low PAPR data transmissions, particularly in the uplink. When $\pi/2$-BPSK symbols undergo spectrum shaping, the PAPR can be reduced to less than $2$ dB. Typically used spectrum shaping filters for $\pi/2$-BPSK transmissions are either a 2-tap filter with an impulse response of $[1~ -1]$ or a 3-tap filter with an impulse response of $[-0.28~ 1~-0.28]$ (these filters will be power normalized to 1 before DFT precoding). Studies, such as~\cite{Ali_txrx, ali_rs_design, IITH_tdoc, IITH_tdoc_2, Kiransir_pi2BPSKpaper}, have demonstrated that $\pi/2$-BPSK can be spectrum-shaped without requiring excess bandwidth, meaning that $\pi/2$-BPSK data transmissions do not need any additional bandwidth, yet achieving nearly full power at PA saturation level.

	In $5$G-NR, unlike other modulation schemes such as QPSK and QAM, which use Zadoff-chu (ZC) sequences for RS. The $\pi/2$-BPSK data transmissions employ $\pi/2$-BPSK RS to maintain identical PAPR for both data and RS transmissions. However, since $\pi/2$-BPSK RS are not spectrally flat sequences, their use with LS estimation leads to performance losses. As shown in Fig.~\ref{fig: 2tapfil}, the frequency domain response of $2$-tap and $3$-tap filters approach zero value at the band edges, which amplifies noise at the band edges, specifically when the LS method is employed. This necessitates additional processing at the receiver to mitigate these losses. To mitigate this issue, in our design, we introduce a regularization factor in~\eqref{eq: LS estimation}, denoted as $\lambda$. 
	\begin{figure}
		\centering
		\includegraphics[width = \linewidth]{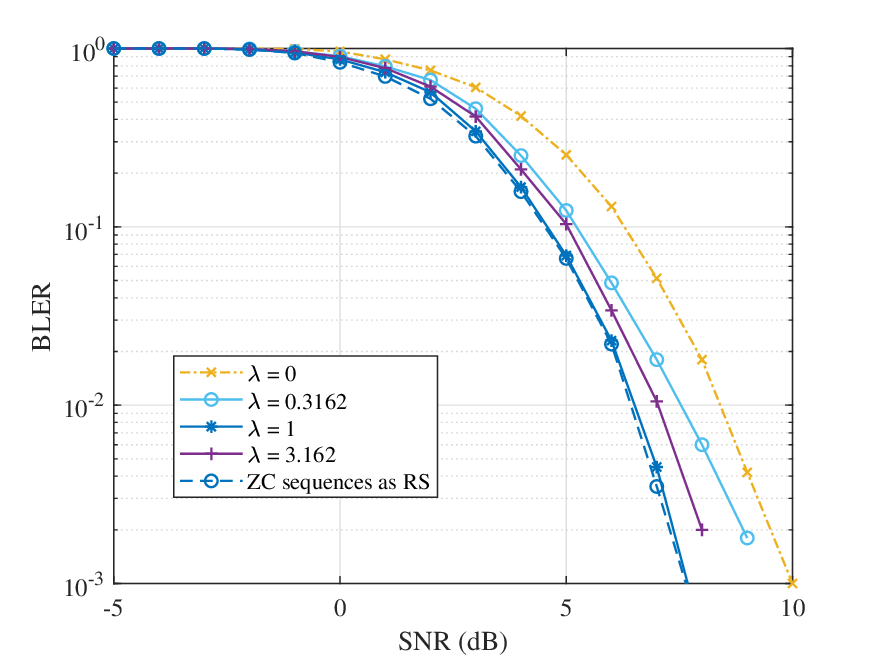}
		\caption{BLER of OTFDM for $\pi/2$-BPSK data multiplexed with $\pi/2$-BPSK RS with different regularization factors for channel estimation in TDL-C $1\mu s$ channel with a carrier frequency of 7GHz, SCS of 30KHz.}
		\label{fig: pi2BPSK_diff_reg_fac}
	\end{figure}

	\begin{equation}
		{h}_{f, {\scalebox{0.5} {${RS}$}}}(k) = {y}_{f, {\scalebox{0.5} {${RS}$}}}(k) \frac {{r}_{f}(k)^*} {|{r}_{f}(k)|^{2} + \lambda},
	\end{equation}
	
	To evaluate the impact of $\lambda$ on BLER performance, link-level simulations were conducted, with the results presented in Fig.~\ref{fig: pi2BPSK_diff_reg_fac}. The figure demonstrates that adding a regularization factor improves decoding performance compared to estimation without any regularization. To determine the optimal value of $\lambda$, we tested values of $0.3162$, $1$, and $3.162$, chosen based on the operating SNRs of $\pi/2$-BPSK modulation. The results indicate that $0.3162$ and $3.162$ have similar performances, while $1$ shows performance similar to ZC sequences as RS.
	
	\subsection{Phase compensation using ARS}\label{subsec: ARS}
	As outlined in Section~\ref{sec: transmitter}, ARS are transmitted alongside regular RS to track phase changes within an OTFDM symbol. High user mobility can cause the OTFDM signal to experience phase variations post-equalization, often in the form of a fixed frequency offset. To mitigate these phase variations, estimating and correcting the frequency offset is necessary. This estimation is performed using the RS and ARS. The equalizer output contains both data and ARS shown in Fig.~\ref{fig: symbol_w_pts}. Let:
	\begin{equation}
		\hat{x}_t (n) = \frac{1}{{M}} \sum_{k = 0}^{M-1} \hat{x}_f (k) e^{\frac{j2\pi kn}{M}}
	\end{equation}
	\begin{equation}\label{eq:RxIDFTsymb}
		\mathbf{\hat{x}}_t = \begin{bmatrix}
			\mathbf{\hat{r}}_{t, {\scalebox{0.5} {${CP}$}}} \\ \mathbf{\hat{r}}_{t} \\ \mathbf{\hat{r}}_{t, { {\scalebox{0.5} {${CS}$}}}} \\ \mathbf{\hat{d}}_{t} \\ \mathbf{\hat{a}}_{t}
		\end{bmatrix}.
	\end{equation}
	Considering that $\theta$ represents the phase growth between successive samples, the resultant equalized ARS samples $\hat{a}_t(n)$ can be represented as follows. 
	\begin{equation}
		\hat{a}_t(n) = a_t(n) e^{j\theta (n+l_{\scalebox{0.5}{CS}}+l_d)}
	\end{equation}
	Here, $a_t(n)$ are the transmitted ARS samples. Hence, the phase estimation can be performed using $a_t(n)$  shown below
	\begin{equation}
		\hat{\theta} = \frac{1}{l_{\scalebox{0.5}{ARS}}}\sum_{n = 0}^{l_{{\scalebox{0.5} {${ARS}$}}-1} }  angle\Bigg( \hat{a}_t(n)\times (a_t(n))^*\Bigg)
	\end{equation}
	Phase growth correction on equalized data samples is as shown below:
	\begin{equation}
		\hat{d}_t(n) = \hat{d}_t (n) e^{ (n+l_{\scalebox{0.5}{CS}})\hat{\theta}}; n = \{1, 2, \dots, l_d\}
	\end{equation}
	
	\begin{table}[t]
		\centering
			{
				\caption{Simulation settings.}
				\label{tab: Sim ParamsFR1}
				\renewcommand{\arraystretch}{1.5}
				{\fontsize{7}{7}\selectfont
					\begin{tabular}{|c|c|c|c|}
						\hline
						\multirow{1}{*} {Parameter} & \multicolumn{3}{c|}{value} \\
						\cline{1-4}
						~~~~Carrier frequency (GHz)~~~~& \multicolumn{3}{c|}{7} \\
						\hline
						System bandwidth (MHz) &~~~100~~~&~~~200~~~&400 \\
						\hline
						Subcarrier spacing (KHz) &~~~30~~~&~~~60~~~&120\\
						\cline{1-4}
						\makecell{Excess bandwidth\\(\% of bandwidth allocation)} & \multicolumn{3}{c|}{$5$}\\
						\hline
						\multirow{1}{*}{Number of symbols} &  \multicolumn{3}{c|}{$1$} \\
						\hline
						\multirow{1}{*}{Modulation (Coding rate)} &  \multicolumn{3}{c|}{\makecell{$\frac{\pi}{2}-BPSK (0.3066) $, QPSK (0.4385), \\$16$-QAM (0.4785),\\ $64$-QAM (0.8) 
								, $256$-QAM (0.894)}} \\
						\hline
						\multirow{1}{*}{Channel Model} &   \multicolumn{3}{c|}{TDL-C $1\mu s$, TDL-C $260$nsec, HST}  \\
						\hline
						\makecell{\multirow{1}{*}{Number of}\\\multirow{1}{*}{UE transmitter antennas}} &  \multicolumn{3}{c|}{$1$}\\
						\hline      
						\makecell{\multirow{1}{*}{Number of }\\\multirow{1}{*}{ BS receiver antennas}} &  \multicolumn{3}{c|}{$1$}\\
						\hline
						\multirow{1}{*}{DMRS sequences} &  \multicolumn{3}{c|}{\makecell{$\pi/2$-BPSK (for $\pi/2$-BPSK data),\\ ZC (for other modulation schemes) }}\\
						\hline
						\multirow{1}{*}{ARS sequences} &  \multicolumn{3}{c|}{\makecell{$\pi/2$-BPSK (for $\pi/2$-BPSK data),\\ ZC (for other modulation schemes) }}\\
						\hline
					\end{tabular}
				}
			}
			
		\end{table}
		
		\begin{figure}[t]
			\centering
			\subfloat[With $8$\% RS overhead, different extension factors\label{fig: diffExtfac}]{%
				\includegraphics[width = \linewidth]{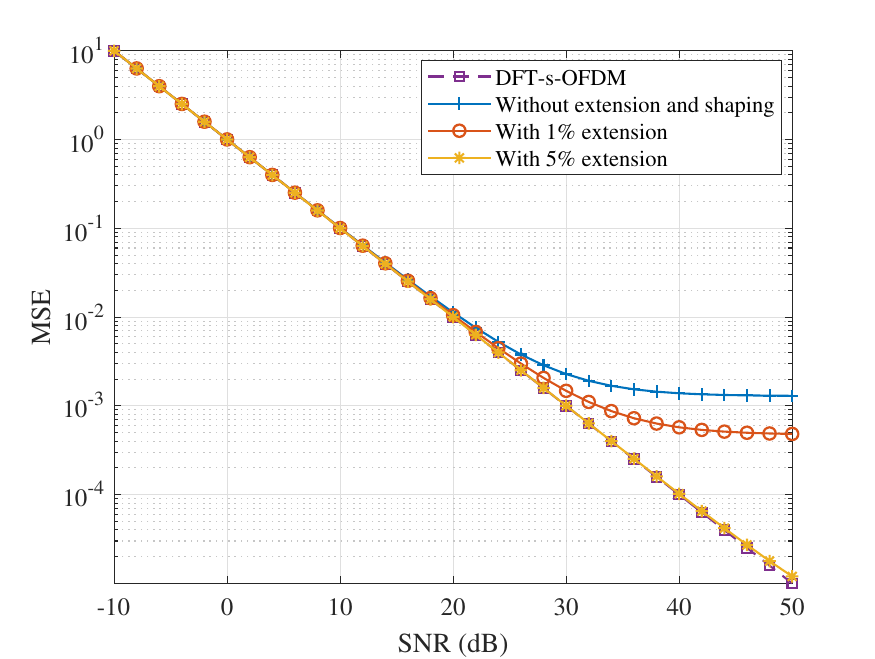}}
			\vfill
			\subfloat[With $5$\% extension factor, different RS overhead\label{fig: diffRSOH}]{%
				\includegraphics[width = \linewidth]{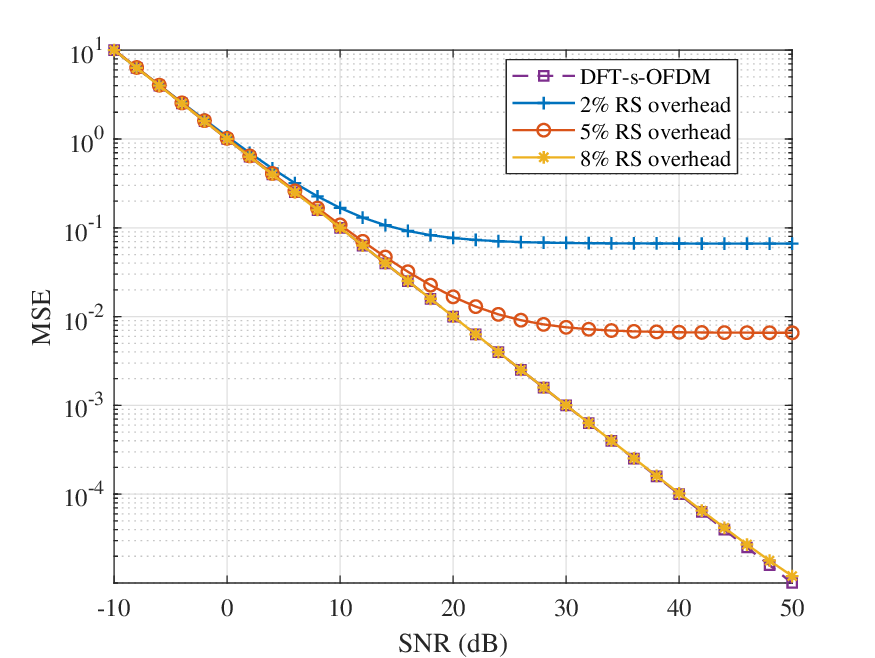}}
			\caption{Mean Squared Error (MSE) between the estimated channel and actual channel acted on OTFDM symbol with different extension factors and RS overheads in TDL-C $1\mu s$ channel.}
			\label{fig: MSE} 
		\end{figure}
		\section{Results}\label{sec: results}

		In this section, we present a comparison of the performance of the proposed OTFDM waveform against the current state-of-the-art. We focus on two primary performance metrics: Block Error Rate (BLER) versus SNR and RS overhead. The performance analysis is carried out across two distinct simulation scenarios: 
		\begin{enumerate}
			\item A dense urban macro environment, where user speeds are generally low. 
			\item A typical rural setting characterized by high user speeds.
		\end{enumerate}  
		For the urban macro scenario, a TDL-C channel with a maximum delay spread of $1 \mu s$ is utilized. In contrast, the rural high-mobility case examines two-channel models: a TDL-C channel with a maximum delay spread of 260 ns and a High-Speed Train (HST) model. Both scenarios use a center frequency of 7 GHz, which is proposed as a potential operating frequency for next-generation technologies~\cite{6gfreq}. 
		All parameters considered for the analysis are outlined in Table~\ref{tab: Sim ParamsFR1}, predominantly obtained from~\cite{PaperChannelmodel}.
		
		\begin{figure}
			\centering
			\includegraphics[width = \linewidth]{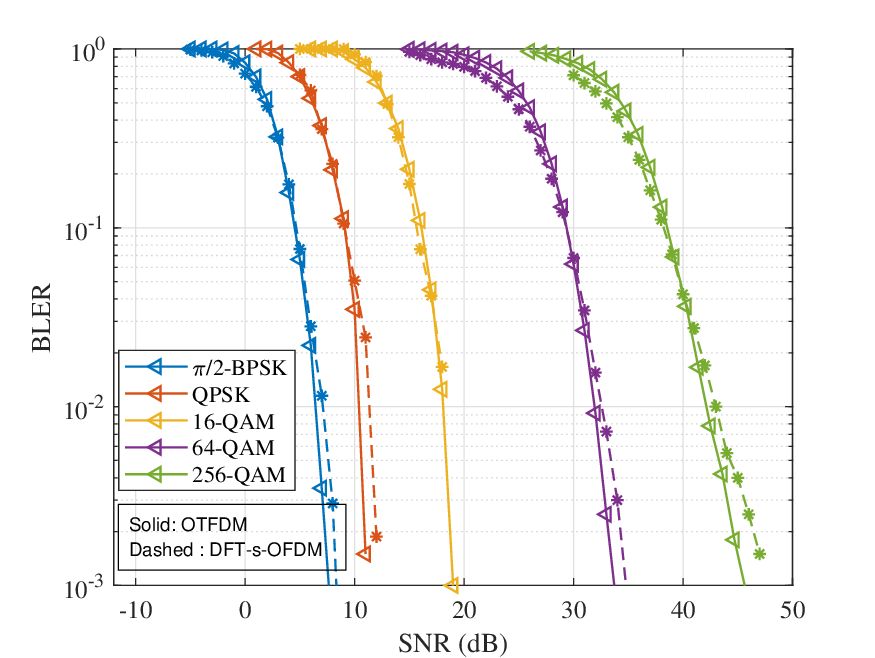}
			\caption{BLER performance comparison of OTFDM with DFT-s-OFDM for different modulations at a carrier frequency of $7$GHz, SCS of $30$KHz in TDL-C $1\mu s$ channel with user speed of $3$ Km/h.}
			\label{fig: 7ghzTDLC100}
		\end{figure}
		\begin{table}[t]
			\centering
				{
					\caption{{Modulation specific RS overheads, extension factors for TDL-C channel with $30$ KHz SCS.}}
					\label{Tab: optimalRS}
					\renewcommand{\arraystretch}{2}
					{\fontsize{6.5}{6.5}\selectfont
						\begin{tabular}{|c|c|c|c|c|c|c|c|} \hline 
							\thead{Modulation} & \thead{RS \\size\\($l_r$)} & \thead{CP \\size}& \thead{CP \\size}&\thead{RS+CP\\+CS \\(\%)} &\thead{Window \\length\\ ($l_n$)}&\thead{Extension\\ factor \\(\%)}& \thead{Over-\\head\\(\%)}\\ 
							\hline \hline
							{$\pi/2$-BPSK} & $72$ & $56$ & $18$ & $4.9$ & $36$ & 0 & $4.9$ \\
							\hline 
							QPSK & $84$ & $63$ & $21$ & $5.4$& $63$& 0 & $5.4$\\
							\hline 
							$16$-QAM & $108$& $81$& $27$ & $7$& $108$& 0 & $7$\\
							\hline 
							$64$-QAM & $120$ & $90$ & $30$ & $7.7$ & $120$ & $5$&$12.7$\\
							\hline
							$256$-QAM & $132$ & $108$ & $34$ & $8.5$ & $132$ & $5$&$13.5$\\
							\hline
						\end{tabular}
					}
				}
			\end{table}
			\vspace{-7.5pt}
			\subsection{BLER performance}\label{sec: blertdl}
			Conventional DFT-s-OFDM systems require at least two symbols: one for data transmission and another for RS transmission. For BLER analysis, a two-symbol configuration is used for the DFT-s-OFDM waveform, while a single-symbol is considered for the proposed OTFDM waveform.
			
			\begin{figure*}[t]
				\centering
				\subfloat[$\pi/2$-BPSK\label{fig: Pi2bpsk_HST_HIGHdOPPLER_14}]{%
					\includegraphics[width = 0.33\linewidth]{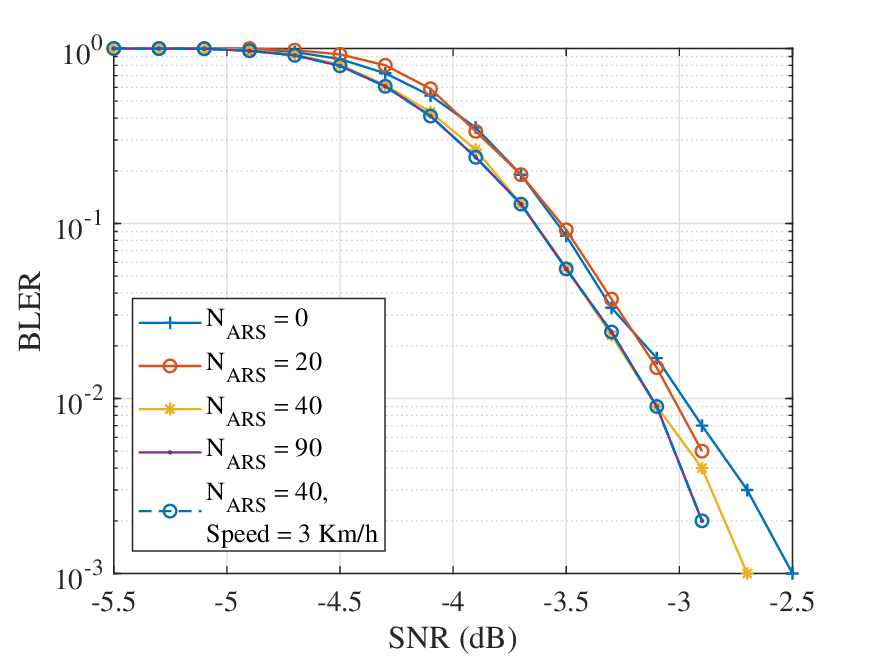}}
				\hfill
				\subfloat[QPSK\label{fig: QPSK_HST_HIGHdOPPLER}]{%
					\includegraphics[width = 0.33\linewidth]{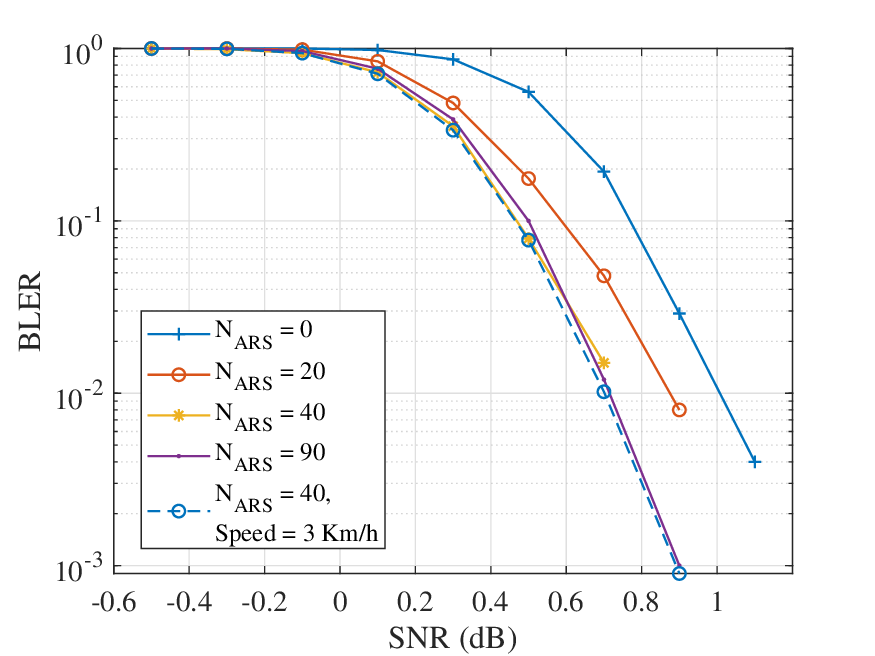}}
				\hfill
				\subfloat[16-QAM\label{fig: 16QAM_HST_HIGHdOPPLER}]{%
					\includegraphics[width=0.33\linewidth]{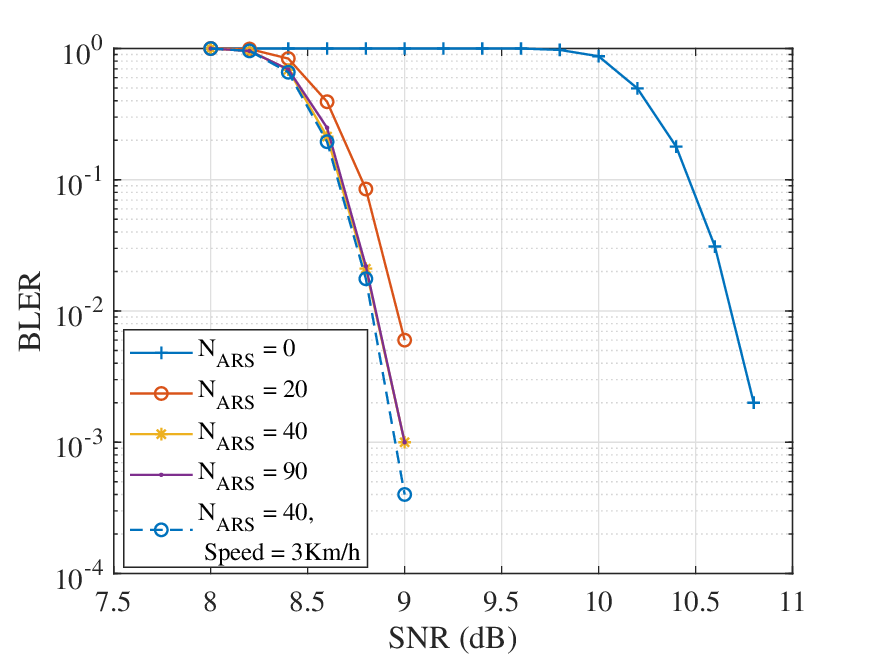}}
				\hfill
				\subfloat[64-QAM\label{fig: 64QAM_HST_HIGHdOPPLER}]{%
					\includegraphics[width=0.33\linewidth]{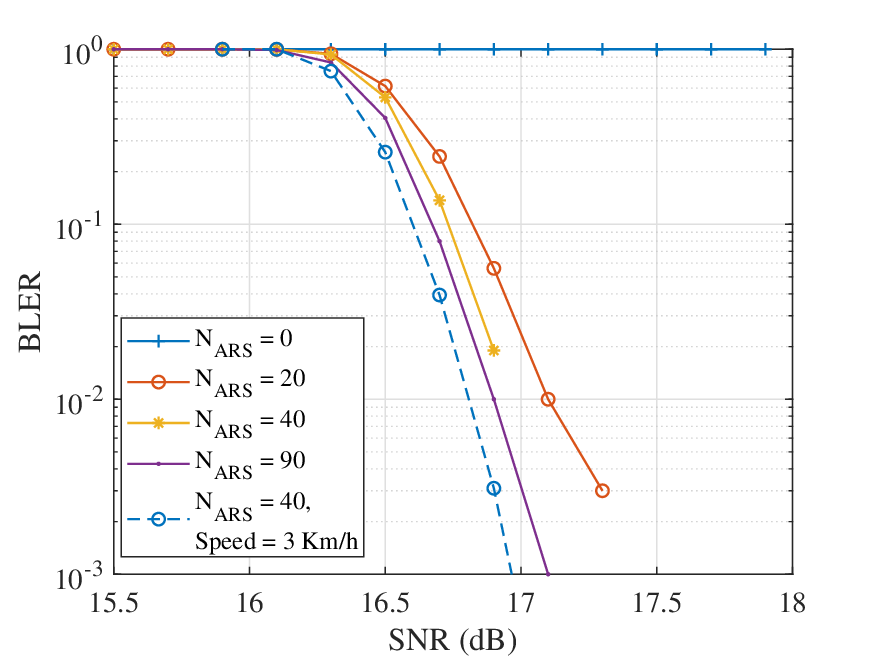}}
				\subfloat[256-QAM\label{fig: 256QAM_HST_HIGHdOPPLER}]{%
					\includegraphics[width=0.33\linewidth]{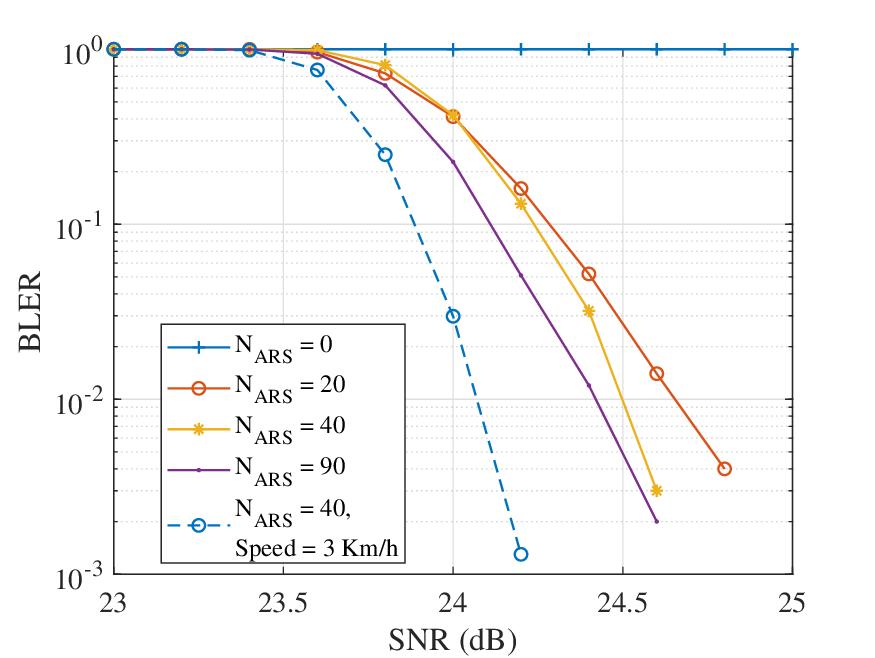}}
				\hfill
				\caption{BLER of OTFDM with different ARS sizes at $500$ Km/h user speed in High-Speed Train (HST) channel model with a carrier frequency of $7$GHz, and SCS of $30$KHz.}
				\label{fig: HST_500} 
			\end{figure*}

			We conducted comprehensive simulations to determine the optimal RS length and excess bandwidth for each modulation order. Fig.~\ref{fig: diffExtfac} shows the Mean Squared Error (MSE) between actual and estimated channels for the proposed system across different extension factors while maintaining a consistent RS overhead. Concurrently, Fig.~\ref{fig: diffRSOH} illustrates the MSE performance for various RS overheads with a constant extension factor. Our findings indicate that an approximately $8\%$ RS overhead combined with a $5\%$ extension factor yields performance metrics similar to the DFT-s-OFDM system. Further, the quality of channel estimates required for successful block decoding varies with the modulation order. Consequently, the RS length and extension factors needed to achieve the desired channel estimation accuracy also vary with modulation size. For the proposed method, the appropriate amount of RS overhead and extension factor for each modulation order is determined empirically through extensive BLER simulations. These simulations identify the optimal RS overhead and extension factor for each modulation order. The modulation specific
			RS overhead and extension factors are tabulated in Table~\ref{Tab: optimalRS}. As shown in the table, lower-order modulations, such as $\pi/2$-BPSK and QPSK, require low RS overhead. However, the RS overhead increases with the modulation size for higher-order modulations. Furthermore, no excess bandwidth is needed for $\pi/2$-BPSK, QPSK, and $16$-QAM modulations, which can tolerate higher estimation errors. For $64$-QAM and $256$-QAM, $5$\% excess bandwidth is considered for the BLER analysis.

			The BLER performance comparison between OTFDM and the conventional DFT-s-OFDM is shown in Fig.~\ref{fig: 7ghzTDLC100}. The existing DFT-s-OFDM systems leverage the advantage of readily available separate full symbol DMRS, thereby enabling a straightforward implementation of the frequency domain LS estimation at the receiver. For the proposed OTFDM waveform, the estimation method discussed in Section~\ref{subsec: Chan esti} is used to obtain the channel estimates on each subcarrier to decode the data. From the figure, it can be noticed that the RS lengths determined empirically are sufficient enough to offer the block error rate performance as DFT-s-OFDM systems. 
			\vspace{-10pt}
			\subsection{Support of OTFDM for high-speed user}\label{sec: blertdl30}
			In this subsection, we present the BLER performance of the OTFDM waveform in scenarios involving very high user speeds. The BLER analysis is conducted using two-channel models. The first is a TDL-C channel model with a maximum delay spread of 260 ns, typically used to simulate rural settings or when a user is moving along highways, where the multipath effects are minimal. The second is a specialized HST model, defined in~\cite{3gpp.38.104}, which is used to analyze performance at extremely high user speeds, such as those encountered by high-speed trains or aircraft. The parameters considered for this analysis are listed in Table~{\ref{tab: Sim ParamsFR1}}. \\
			\begin{table}[t]
				\centering
					{
						\caption{{Recommended speed in Km/h for a given SCS and modulation in TDL-C channel with a maximum delay spread of 260 ns.}}
						\label{Tab: Maxspeed}
						\renewcommand{\arraystretch}{2}
						{\fontsize{7}{7}\selectfont
							\begin{tabular}{|c|c|c|c|c|c|} \hline 
								\thead{Modulation} & \thead{$\pi/2$-BPSK} & \thead{QPSK}&\thead{$16$-QAM}& \thead{$64$-QAM}&\thead{$256$-QAM} \\ 
								\hline \hline
								{$30$ KHz} & $> 500$ & $> 500$ & $> 500$& $100$ & $30$\\
								\hline 
								{$60$ KHz} & $>500$ & $>500$ & $>500$& $160$ & $70$\\
								\hline 
								{$120$ KHz} & $>500$ & $>500$ & $>500$& $300$ & $100$\\
								\hline 
							\end{tabular}
						}
					}
				\end{table}
				
				\subsubsection{BLER for high Doppler in TDL-C channel}
				
				In this subsection, we present the BLER performance of the OTFDM waveform for different modulation schemes and user speeds using the TDL-C channel model. To determine the maximum user speeds supported by the proposed design and the conventional DFT-s-OFDM waveform, we examined system performance at various user speeds with an SCS of $30$ kHz. The maximum speed supported by each waveform for various modulation schemes is shown in Table~\ref{Tab: Maxspeed}, with the corresponding BLER performance plots in Fig.~\ref{fig: HIGHSPEED_ALLMOD_TDLC_30KHz} of Appendix~\ref{app: BLERhighspeed}.
				\begin{figure*}[t]
					\centering
					\subfloat[DFT-s-OFDM\label{fig: papr_DFT-s-OFDM}]{%
						\includegraphics[width = 0.33\linewidth]{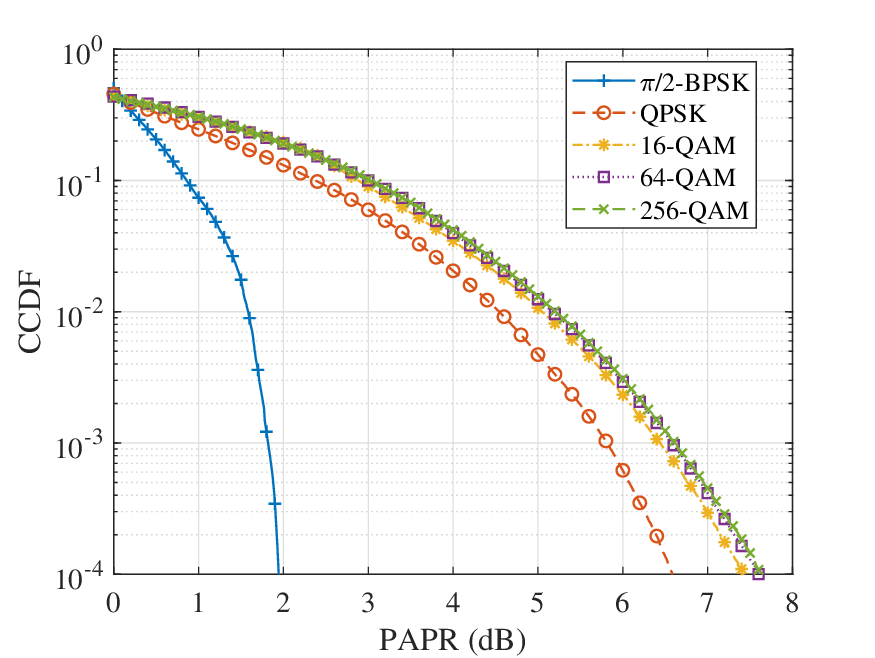}}
					\hfill
					\subfloat[OTFDM with $5$\% extension\label{fig: 5_per_ext}]{%
						\includegraphics[width=0.33\linewidth]{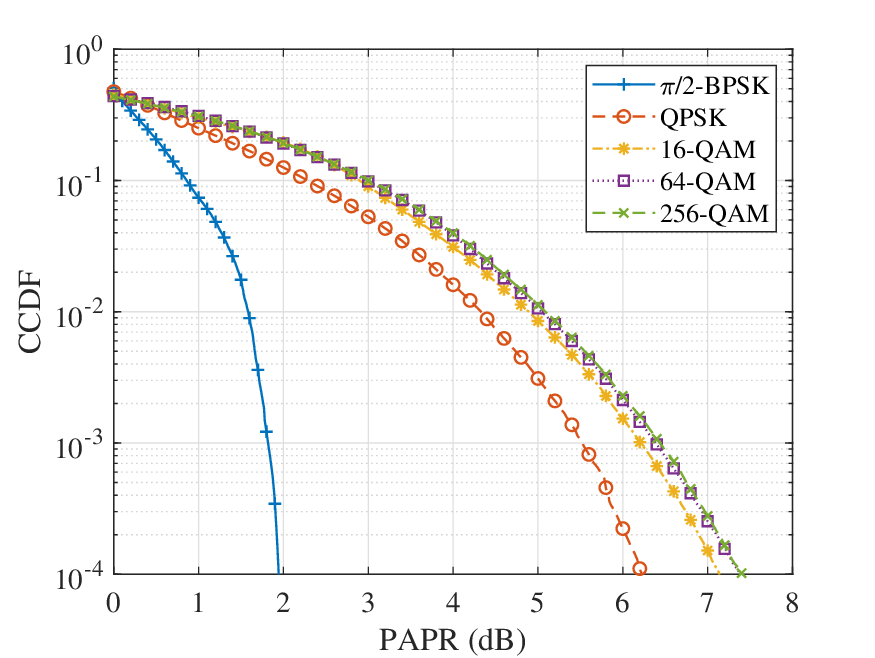}}
					\hfill
					\subfloat[OTFDM with $10$\% extension\label{fig: 10_per_ext}]{%
						\includegraphics[width=0.33\linewidth]{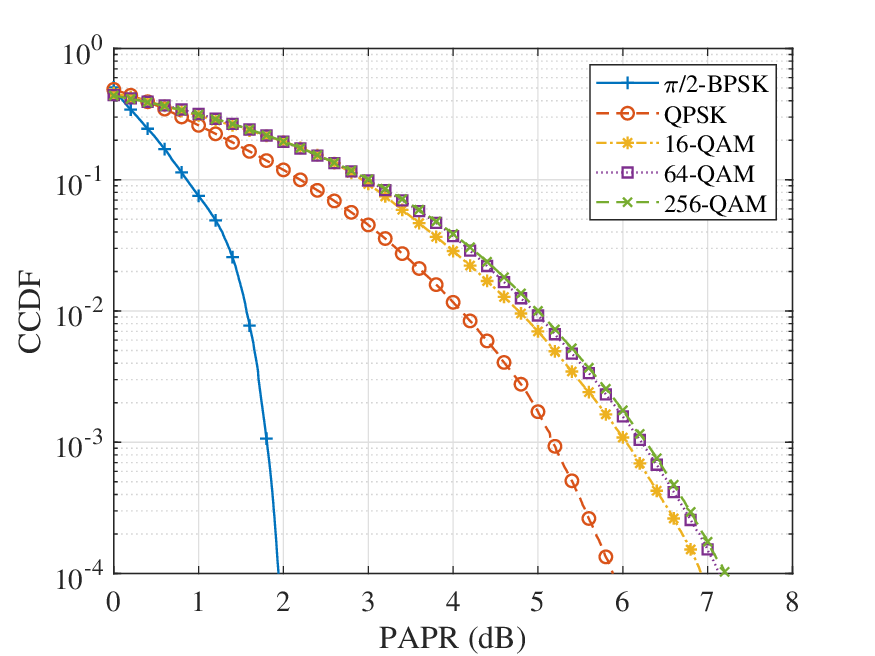}}
					\hfill
					\caption{PAPR DFT-s-OFDM and the OTFDM waveforms with different excess bandwidth values.}
					\label{PAPR_diff_ext} 
				\end{figure*}
				
				For OTFDM, it can be observed from Fig.~\ref{fig: HIGHSPEED_ALLMOD_TDLC_30KHz} and Table~\ref{Tab: Maxspeed} that the waveform can support $\pi/2$-BPSK, QPSK, and 16-QAM for user speeds up to $500$ Km/h with only marginal performance loss. Notably, lower-order modulations are more resilient to temporal variations in the received signal, allowing them to be supported even at higher user speeds. However, higher-order modulation schemes such as $64$-QAM and $256$-QAM are more susceptible to channel phase errors. Despite this, it can be seen from the figures that the proposed OTFDM waveform offers better support at higher user speeds compared to the existing DFT-s-OFDM waveform across all modulation schemes.
				
\begin{figure}[htb]
					\centering
					\subfloat[BLER performance with $\pi/2$-BPSK, QPSK and $16$-QAM modulation schemes\label{fig: pi2BPSK_tdlC_HIGHdOPPLER60}]{%
						\includegraphics[width = 0.5\linewidth]{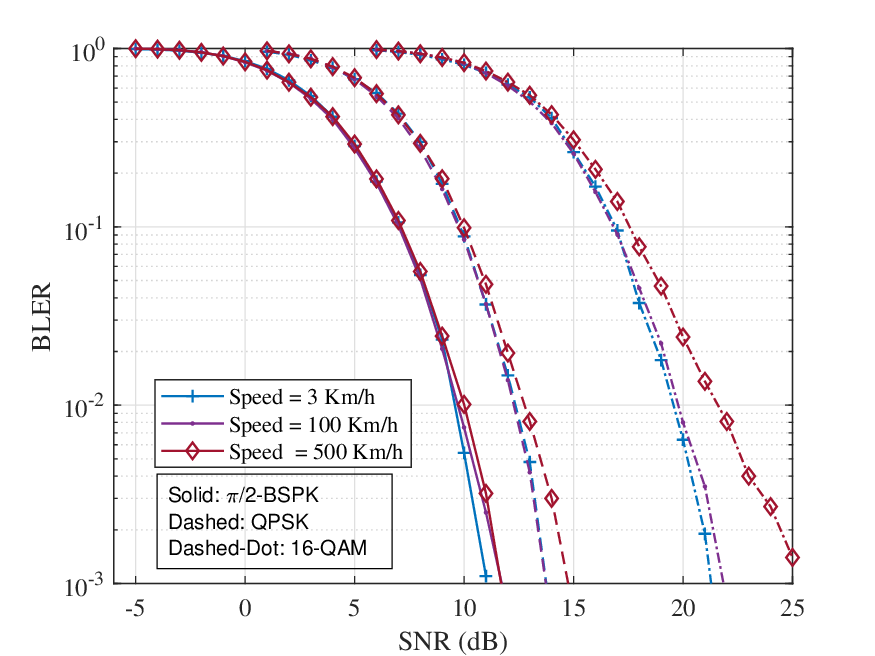}}
					\subfloat[BLER performance with $64$-QAM and $256$-QAM modulation schemes\label{fig: QPSK_tdlc_HIGHdOPPLER60}]{%
						\includegraphics[width = 0.5\linewidth]{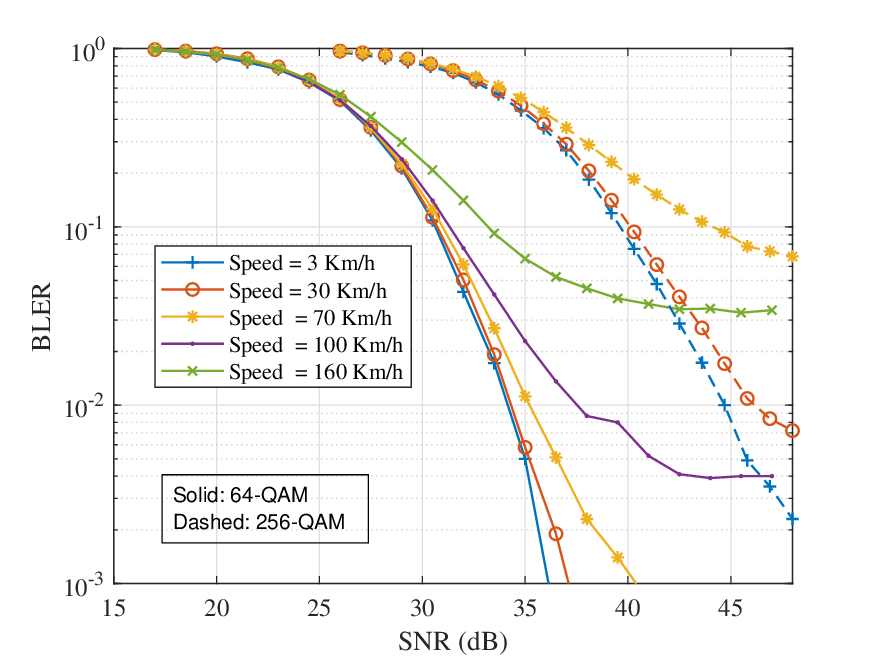}}
					\hfill
					\caption{BLER of OTFDM for different modulation schemes and user speeds in TDL-C $260$ ns channel with the carrier frequency of $7$GHz, and SCS of $60$KHz.}
					\label{fig: HIGHSPEED_ALLMOD_TDLC_60KHz} 
				\end{figure}

\begin{figure}[htb]
					\centering
					\subfloat[BLER performance with $\pi/2$-BPSK, QPSK and $16$-QAM modulation schemes\label{fig: pi2BPSK_tdlC_HIGHdOPPLER120}]{%
						\includegraphics[width = 0.5\linewidth]{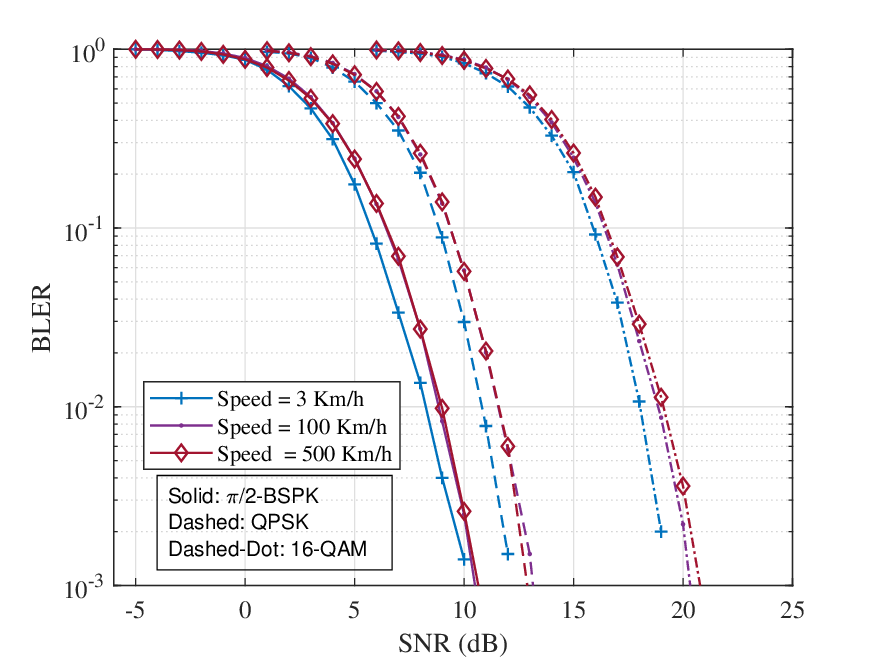}}
					\subfloat[BLER performance with $64$-QAM and $256$-QAM modulation schemes\label{fig: QPSK_tdlc_HIGHdOPPLER120}]{%
						\includegraphics[width = 0.5\linewidth]{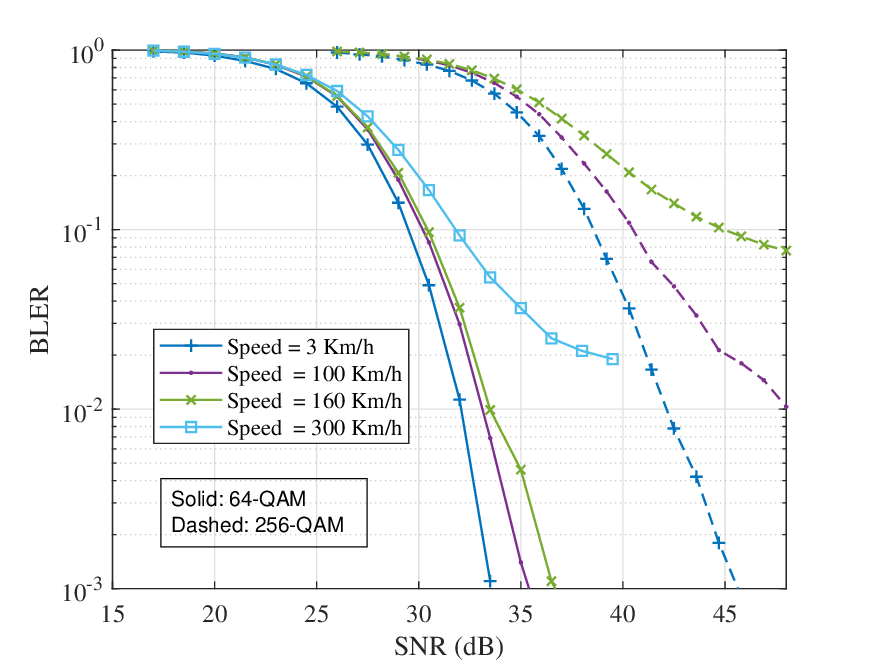}}
					\hfill
					\caption{BLER of OTFDM for different modulation schemes and user speeds in TDL-C $260$nsec channel with the carrier frequency of $7$GHz, and SCS of $120$KHz.}
					\label{fig: HIGHSPEED_ALLMOD_TDLC_120KHz} 
				\end{figure}

				The extent of Doppler shift resulting from user mobility depends on the SCS. Therefore, system performance is evaluated for SCS values of $60$ kHz and $120$ kHz, with varying user speeds. Appendix~\ref{app: BLERhighspeed} includes Figs.~\ref{fig: HIGHSPEED_ALLMOD_TDLC_60KHz}, \ref{fig: HIGHSPEED_ALLMOD_TDLC_120KHz} that illustrate the BLER performance. Table~\ref{Tab: Maxspeed} lists the maximum user speeds supported by the OTFDM waveform for each combination of SCS and modulation order in the TDL-C channel. For a $120$ kHz SCS, both QPSK and $16$-QAM support speeds greater than $500$ Km/h, while $64$-QAM and $256$-QAM support speeds up to $300$ Km/h and $160$ Km/h, respectively. Typically, user speeds in general-purpose use cases are below $150$ Km/h, where OTFDM performs well. However, very high user speeds, up to $500$ Km/h, as considered by IMT-$2030$, may be relevant for high-speed trains or aircraft. For these high-speed scenarios, a specialized channel model, mainly the High-Speed Train (HST) model, is defined in~\cite{3gpp.38.104} to evaluate the link performance.
				
				\subsubsection{Enhancing OTFDM performance using ARS in high-speed use cases}
				In this subsection, we show that the performance of OTFDM can be enhanced to support very high user speeds by employing ARS. As explained in Section~\ref{subsec: ARS}, ARS are used to mitigate phase variations caused by high user mobility. By integrating ARS with RS, the OTFDM waveform can support speeds up to 500 Km/h, even for higher-order modulations. The required length of ARS to achieve the desired system performance is determined empirically through extensive block error simulations. we have considered the HST channel model, assuming $D_{s}$ to be $300$m and $D_{min}$ to be $2$m, defined in~\cite{3gpp.38.104} for high-speed scenarios.

				Fig.~\ref{fig: HST_500} shows the BLER performance of the OTFDM waveform with different modulation schemes when transmitted over the HST channel model at a speed of 500 Km/h. Since the estimation accuracy of the phase (resulting from the Doppler) depends on the size of ARS used for estimation, different sizes of ARS are considered for the evaluation of each modulation scheme. It can be observed from the figure that in addition to RS, ARS of length 1-2\% of the allocated resources are needed for the phase correction to support speeds up to 500 Km/h without BLER performance degradation. Moreover, higher SCS can be used for HST to support speeds beyond 500 Km/h, even for higher-order modulations.
				\vspace{-7.5pt}
				\subsection{PAPR performance}
				Fig.~\ref{PAPR_diff_ext} shows the Complementary Cumulative Distribution Function (CCDF) of the PAPR for the proposed waveform with excess bandwidth of $5$\% and $10$\%. The PAPR performance is compared to the conventional DFT-s-OFDM waveform across various modulation schemes. As discussed in section~\ref{Subsec: pi2data}, no excess bandwidth is required for the OTFDM symbol when operating with  $\pi/2$-BPSK modulation. For other modulation schemes, excess bandwidth of $5$\% and $10$\% are analyzed. The figures indicate that the proposed system achieves a PAPR gain of approximately $0.5$ dB for QPSK modulation compared to the DFT-s-OFDM waveform. For higher-order modulations, the PAPR remains similar to the DFT-s-OFDM waveform, showing no degradation, as shown in Table.~\ref{tab: PAPR gain}. Hence, the proposed waveform can be operated at PA saturation point for $\pi/2$-BPSK modulation, and for other modulations, PA back-off defined in~\cite{3gpp.38.101-1} can be reused. The low PAPR OTFDM output signal becomes useful for DPD for high-power transmissions. Further, it is noteworthy that the PA back-off required for OTFDM is substantially lower than CP-OFDM defined in~\cite{3gpp.38.101-1}. Additionally, since the PAPR gain for OTFDM depends on the extension factor used, increasing the extension factor and shaping can enhance the gain further.
				\begin{table}[t]
					\centering
					{
						\caption{PAPR gain of OTFDM compared to the DFT-s-OFDM.}
						\label{tab: PAPR gain}
						\renewcommand{\arraystretch}{1.5}
						\fontsize{7}{7}\selectfont
						\begin{tabular}{|c|c|c|}
							\hline
							\multirow{2}{*}{Modulation} &
							\multicolumn{2}{|c|}{Gain at $1\%$ CCDF point} \\
							\cline{2-3}
							& $5\%$ extension & $10\%$ extension \\
							\hline
							QPSK & $0.21$ & $0.47$\\
							\hline
							$16$ QAM & $0.16$ & $0.28$\\
							\hline
							$64$ QAM & $0.13$ & $0.23$ \\
							\hline
							$256$ QAM & $0.125$ & $0.21$ \\
							\hline
						\end{tabular}
					}
				\end{table} 
				\vspace{-15pt}
				\subsection{RS overhead}
				In this subsection, we analyze the minimum DMRS overhead required for successful packet detection in both the proposed and conventional DFT-s-OFDM systems. In the conventional DFT-s-OFDM framework, RS and data are transmitted in separate symbols, with up to $4$ symbols dedicated to DMRS transmission. This results in a fixed DMRS overhead of $4/14$ or $28.6$\%, regardless of the modulation order used. In contrast, in the proposed OTFDM framework, the minimum DMRS length is configurable and varies with the type of modulation scheme used for data transmissions. Therefore, the RS length and the extension factor can be dynamically adjusted. The modulation-specific RS overhead and the required extension factors for the OTFDM framework are discussed in detail in section~\ref{sec: blertdl} and are tabulated in Table~\ref{Tab: optimalRS}. It can be observed that the RS overhead required for the OTFDM framework is consistently lower than that of the conventional DFT-s-OFDM framework.

				Considering the ability to time multiplex RS and Data in a single symbol and the ability to support high mobility, low latency, and low PAPR, the waveform is an excellent candidate for use in the upcoming  IMT-$2030$ systems.
				
				\section{Conclusion}
				This paper introduces OTFDM as a candidate waveform for the IMT-2030 framework. OTFDM addresses the limitations of traditional OFDM by enabling simultaneous transmission of RS and data within a single OFDM symbol to support high mobility use cases. OTFDM offers low PAPR transmissions, thereby enhancing PA efficiency. Through multiplexing RS and data in the same symbol followed by excess bandwidth addition and spectrum shaping, OTFDM mitigates challenges related to channel estimation and inter-symbol interference, leading to improved packet error performance. Computer simulations using 3GPP channel models validate the effectiveness of OTFDM in achieving self-contained channel estimation for modulation sizes up to 256 QAM. The simulation results also demonstrate low PAPR of OTFDM. The compatibility of OTFDM with standard OFDM, along with its computationally efficient signal generation and receiver processing using FFT and IFFT, further enhances its practicality and applicability. Overall, with its unique capabilities, OTFDM offers a new waveform paradigm for next-generation mobile communication (6G), addressing extremely low latency requirements, high data rates, and improved power efficiency.
				
				\appendices
				\section{OTFDM symbol for RS with only CP}\label{app: onesided}
\begin{figure}[h]
					\centering
					\subfloat[OTFDM symbol without Additional-RS and RS with only CP.\label{fig: onesided_symbol_wo_pts}]{%
						\includegraphics[width = 0.7\linewidth]{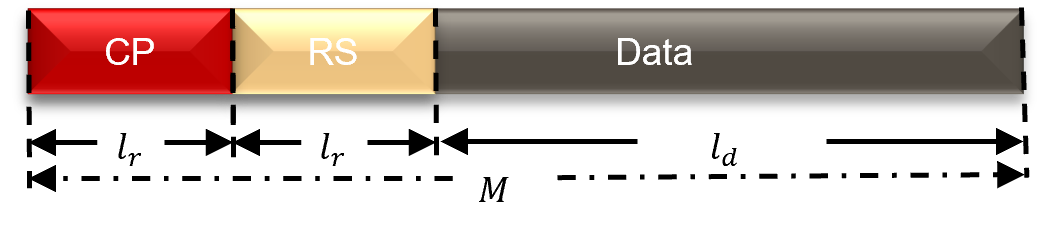}}
					\vfill
					\subfloat[RS extraction for channel estimation.\label{fig: RSsamponesided}]{%
						\includegraphics[width = 0.7\linewidth]{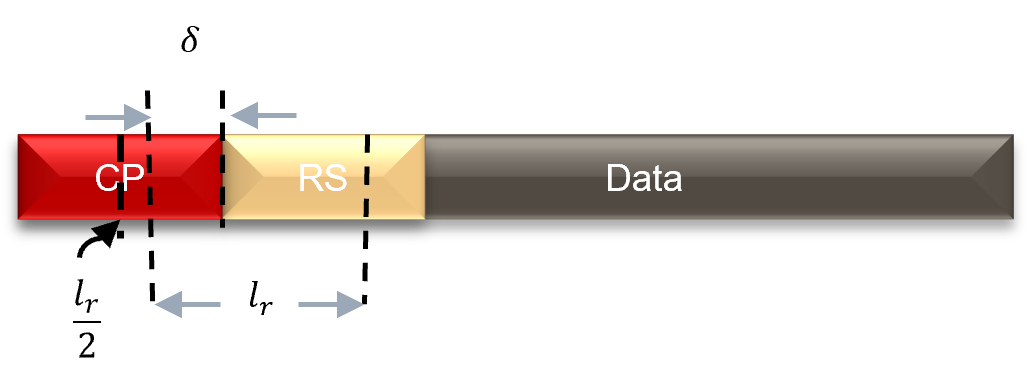}}
					\caption{OTFDM symbol with one-sided CP.}
					\label{fig: onesided_symb_struct} 
				\end{figure}

				{An alternative way for the multiplexed symbol to generate an OTFDM waveform is by appending RS with only CP, as shown in Fig.~\ref{fig: onesided_symbol_wo_pts}. Since RS is not appended with CS, extracting RS from the RS boundary introduces ISI between RS and data. To mitigate this interference, the CP size is kept the same as the RS, and at the receiver, RS is extracted starting from a point inside the CP, as shown in Fig.~\ref{fig: RSsamponesided}. This way, the edge RS samples affected by ISI are neglected, allowing for better channel estimation.}
				
				\section{BLER performance in high-speed scenario}
				\label{app: BLERhighspeed}
				\begin{figure}[htb]
					\vspace{-0pt}
					\centering
					\subfloat[BLER performance with $\pi/2$-BPSK, QPSK and $16$-QAM modulation schemes\label{fig: pi2BPSK_tdlC_HIGHdOPPLER}]{%
						\includegraphics[width = 0.5\linewidth]{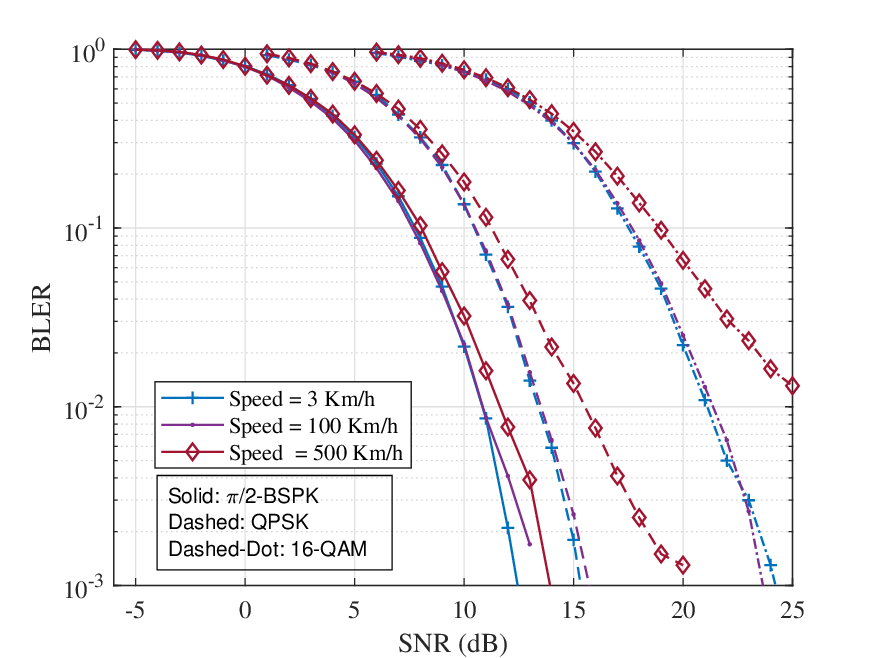}}
					\subfloat[BLER performance with $64$-QAM and $256$-QAM modulation schemes\label{fig: QPSK_tdlc_HIGHdOPPLER}]{%
						\includegraphics[width = 0.5\linewidth]{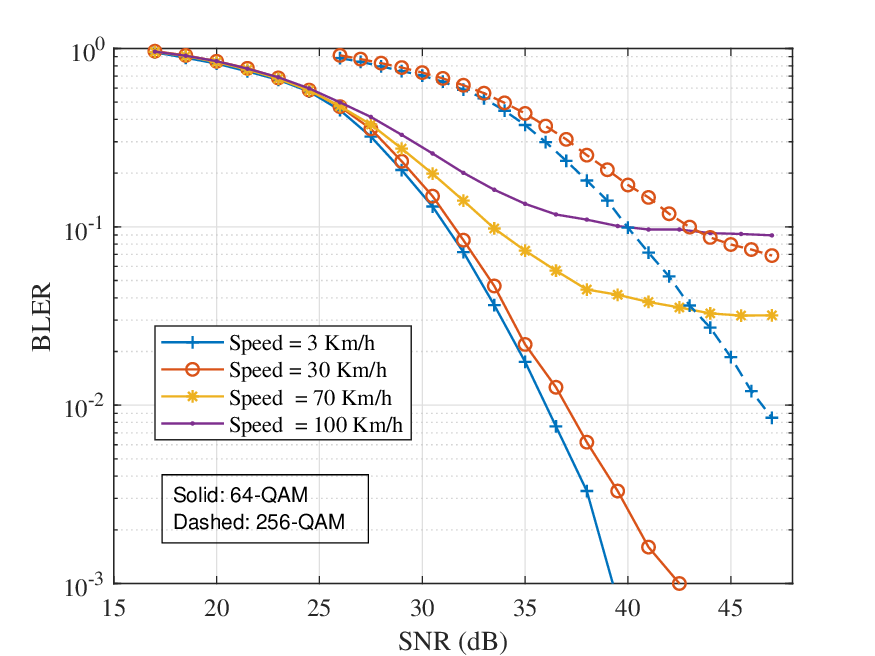}}
					\hfill
					\caption{BLER of OTFDM for different modulation schemes and user speeds in TDL-C $260$nsec channel with the carrier frequency of $7$GHz, and SCS of $30$KHz.}
					\label{fig: HIGHSPEED_ALLMOD_TDLC_30KHz} 
				\end{figure}
				
				\bibliographystyle{ieeetr}
				\bibliography{bibliography_1}
				\vspace{-100pt}
				\begin{IEEEbiography}[{\includegraphics[width=1in,height=1.25in, clip,keepaspectratio]{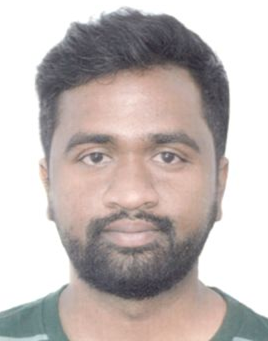}}]{Koteswara Rao Gudimitla} received the B.Tech. degree in electrical and electronics engineering from Sri Venkateswara University College of Engineering, Tirupati, India, in $2015$. He is currently pursuing a Ph.D. degree from IIT Hyderabad, Hyderabad. His research interests include physical-layer algorithms and the transceiver design for next-generation wireless systems.
				\end{IEEEbiography}
				\vspace{-100pt}
				\begin{IEEEbiography}[{\includegraphics[width=1in,height=1.25in, clip,keepaspectratio]{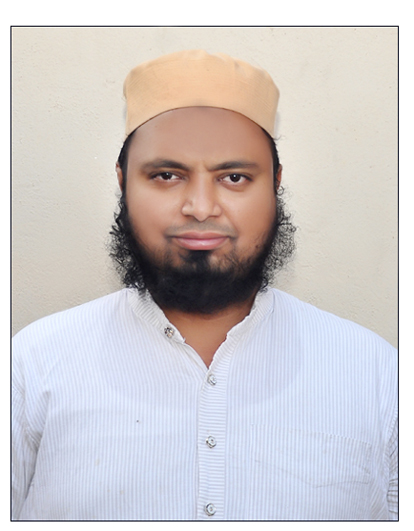}}]{Sibgath Ali khan M} completed his B.Tech degree from JNTU Hyderabad and his Ph.D. in Electrical Engineering from IIT Hyderabad. From 2012 to 2017, he contributed to the Cyber-Physical Systems Lab at IIT Hyderabad, where he focused on the design and real-time implementation of Physical layer algorithms for 4G and 5G systems. In 2018, he joined WiSig Networks and currently serves as a Principal Architect, leading the design and development of algorithms and architectures for Massive MIMO systems. He holds over 15 granted patents and works on next-generation wireless systems.
				\end{IEEEbiography}
				\vspace{-100pt}
				\begin{IEEEbiography}[{\includegraphics[width=1in,height=1.25in, clip,keepaspectratio]{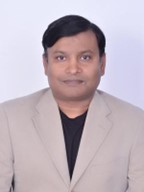}}]{Prof. KIRAN KUCHI} holds a position as a Professor within the Department of Electrical Engineering at the Indian Institute of Technology Hyderabad. He spearheads research and standardization endeavors for 5G-advanced and 6G technologies on behalf of India at international standards organizations focused on shaping future communication systems. With a prolific portfolio, Professor Kuchi boasts over 200 international patents, including some recognized as essential to 5G standards. Notably, he is the founder of WiSig Networks Pvt Ltd, a venture incubated at IITH, where collaborative efforts between IITH and WiSig have led to the development and commercialization of 5G base station and user equipment (UE) technologies, as well as NB-IoT System-on-Chip (SoC) solutions.
					LinkedIn: in/kiran-kuchi-88113b2
				\end{IEEEbiography}

			\end{document}